\documentclass[iop]{emulateapj}
\usepackage{graphicx}
\usepackage{amssymb}
\usepackage{amsmath}
\usepackage{natbib}
\usepackage{url}
\nocite{*}

\submitted{Accepted to ApJ, November 29, 2016}

\shortauthors{TOFFLEMIRE ET AL.}
\begin{document}

\title{Accretion and Magnetic Reconnection in the Classical T Tauri Binary DQ
  Tau}

\author{Benjamin M.\ Tofflemire\altaffilmark{1,7},
Robert D.\ Mathieu\altaffilmark{1,7},
David R.\ Ardila\altaffilmark{2},
Rachel L.\ Akeson\altaffilmark{3},
David R.\ Ciardi\altaffilmark{3},
Christopher Johns-Krull\altaffilmark{4},
Gregory J.\ Herczeg\altaffilmark{5}, \&
Alberto Quijano-Vodniza\altaffilmark{6}}

\altaffiltext{1}{Department of Astronomy, University of Wisconsin--Madison,
  475 North Charter Street, Madison, WI 53706, USA}
\altaffiltext{2}{The Aerospace Corporation, M2-266, El Segundo, CA 90245, USA}
\altaffiltext{3}{NASA Exoplanet Science Institute, IPAC/Caltech, Pasadena, CA
  91125, USA} 
\altaffiltext{4}{Department of Physics and Astronomy, Rice University,
  Houston, TX 77005, USA} 
\altaffiltext{5}{The Kavli Institute for Astronomy and Astrophysics, Peking
  University, Beijing 100871, China} 
\altaffiltext{6}{University of Nari\~{n}o Observatory, Pasto, Nari\~{n}o,
  Colombia} 
\altaffiltext{7}{Visiting astronomer, Kitt Peak National Observatory,
  National Optical Astronomy Observatory, which is operated by the
  Association of Universities for Research in Astronomy (AURA) under a
  cooperative agreement with the National Science Foundation.}

\begin{abstract}

 Binary star-formation theory predicts that close binaries ($a<100$ AU) will
 experience periodic pulsed accretion events as streams of material form at
 the inner edge of a circumbinary disk, cross a dynamically cleared gap, and
 feed circumstellar disks or accrete directly onto the stars. The archetype
 for the pulsed-accretion theory is the eccentric, short-period, classical T
 Tauri binary DQ Tau.  Low-cadence ($\sim$daily) broadband photometry has
 shown brightening events near most periastron passages, just as numerical
 simulations would predict for an eccentric binary. Magnetic reconnection
 events (flares) during the collision of stellar magnetospheres near
 periastron could, however, produce the same periodic, broadband behavior when
 observed at a one-day cadence. To reveal the dominate physical mechanism seen
 in DQ Tau's low-cadence observations, we have obtained continuous,
 moderate-cadence, multi-band photometry over 10 orbital periods, supplemented
 with 27 nights of minute-cadence photometry centered on 4 separate periastron
 passages.  While both accretion and stellar flares are present, the dominant
 timescale and morphology of brightening events are characteristic of
 accretion. On average, the mass accretion rate increases by a factor of 5
 near periastron, in good agreement with recent models. Large variability is
 observed in the morphology and amplitude of accretion events from
 orbit-to-orbit. We argue this is due to the absence of stable circumstellar
 disks around each star, compounded by inhomogeneities at the inner edge of
 the circumbinary disk and within the accretion streams themselves.
 Quasi-periodic apastron accretion events are also observed, which are not
 predicted by binary accretion theory.

\end{abstract}


\section{INTRODUCTION}
\label{intro}

One of the primary outcomes of binary star formation theory is that the
interaction between close binary star systems and their disk(s) is
fundamentally different than the well-established single-star paradigm. In
single stars, interplay between the star and disk is mediated by the stellar
magnetic field \citep{Shuetal1994,Hartmannetal1994}. In this magnetic
accretion model, strong stellar magnetic fields truncate the inner edge of the
disk at the distance where viscous ram pressure balances the magnetic
pressure. This ``magnetospheric radius'' is modeled as 5--10 stellar radii
($\sim$0.05AU; \citealt{Johnstoneetal2014}) for typical pre-MS magnetic field
strengths ($\sim$1--2 kG; \citealt{Johns-Krull2007}) and accretion rates
(10$^{-12}$--$10^{-8}M_\odot$ yr$^{-1}$; \citealt{Alcalaetal2014}). Inside
this radius, material is confined to flow along magnetic field lines where it
impacts the stellar surface at magnetic footpoints, shock-heating the
photosphere \citep{Orlandoetal2013}.

The single-star magnetic accretion model plays a critical role in the
evolution of the star-disk system. For the star, it provides an avenue for
continued mass growth while regulating the stellar angular momentum through
magnetic disk locking \citep{Shuetal1994}. For the disk, accretion processes
set the evolution timescale by controlling the consumption rate, outflow rates
through wind and jet launching, and the intensity of UV radiation relevant for
photoevaporation and disk chemistry \citep{Alexanderetal2014}. By governing
the stability, lifetime, and chemistry of protoplanetary disks, the star-disk
interaction plays a vital role in the formation and evolution planets.

The successes of the single-star accretion paradigm and its impact on the
evolution of the star-disk system highlights the need to characterize the
binary-disk interaction. Most pressing is the indication that binary and
higher multiple systems are a common outcome of star-formation
\citep{Raghavanetal2010}.  \citet{Krausetal2011}, for instance, find that up
to 75\% of Class II/III members of the Taurus-Auriga star-forming region are
in multi-star systems. In binary systems with separations on the order of
typical protostellar disk radii ($\sim$100s of AU; \citealt{Jensenetal1996};
\citealt{Harrisetal2012}) the single-star model cannot simply be applied to
environments where the distribution of disk material and mass flows are more
complex. While theory describing binary-disk interaction is advancing, many of
its predictions remain untested and therefore the effects of binarity on star
and planet formation remains largely unconstrained.

Theory describing the binary-disk interaction in short-period systems has made
two predictions that portray a complex and variable environment compared to
single-stars. First, through co-rotational and Lindblad resonances, orbital
motion will dynamically clear a central region around the binary creating up
to three stable accretion disks: a circumstellar disk around each star and an
encompassing circumbinary disk \citep{Artymowicz&Lubow1994}. Observational
support for this spatial structure has come from modeling the IR spectral
energy distribution (SED) of spectroscopic binaries
\citep{Jensen&Mathieu1997,Bodenetal2009} and from spatially resolving central
gaps from scattered light \citep{Becketal2012} and mm/sub-mm images
\citep{Andrewsetal2011,Harrisetal2012} of longer-period systems.

\begin{deluxetable}{lcc}
\tablewidth{0pt}
\tabletypesize{\footnotesize}
\tablecaption{DQ Tau System Summary}
\tablehead{
  \colhead{Parameter} &
  \colhead{Value} &
  \colhead{Reference}}
\startdata
P (days)                         & $15.80158 \pm 0.00066$ & 1\\
$e$                              & $0.568 \pm 0.013$      & 1\\
T$_{\rm peri}$ (HJD-2,400,000)    & $47433.507 \pm 0.094$  & 1\\
$a$ ($R_\odot$)                  & $28.96 \pm 0.48$       & 1\\
$q\equiv$ M2/M1                      & $0.936 \pm 0.051$     & 1 \\
Periastron Separation ($R_\odot$)& $12.51 \pm  0.43$    & 1 \\
Apastron Separation ($R_\odot$)  & $45.42 \pm  0.43$    & 1 \\
i ($^\circ$)                      & $158 \pm 2$           & 1\\
Rotation Period (d)              & $\sim$2 & 2 \\
Disk $M_{\rm gas}$ ($10^{-4}M_\odot$)           & $>$10 & 3\\
Disk $M_{\rm dust}$ ($10^{-4}M_\odot$)           & 0.90 & 3\\
d (pc)                        & 140 & 4\\
$A_V$                         & $1.5\pm 0.3$  &  5\\

\sidehead{Primary}
$M_1$ ($M_\odot$)       & $0.63 \pm 0.13$    & 1\\
$T_1$ ($K$)             & $3700\pm200$      & 1\\
$L_1$ ($L_\odot$)       & $0.19\pm 0.07$ & 1\\
$R_1$ ($R_\odot$)       & $1.05\pm 0.22$ & 1 \\

\sidehead{Secondary}
$M_2$ ($M_\odot$)       & $0.59 \pm 0.13$    & 1\\
$T_2$ ($K$)                 & $3500\pm175$  & 1\\
$L_2$ ($L_\odot$)        & $0.13\pm0.07$     & 1\\
$R_2$ ($R_\odot$)         & $1.00\pm 0.21$   & 1 \\

\enddata 
\tablecomments{$^{(1)}$\citep{Czekalaetal2016},
  $^{(2)}$\citep{Basrietal1997}, $^{(3)}$\citep{Williams&Best2014},
  $^{(4)}$\citep{Kenyonetal1994}, $^{(5)}$(This work)}
\label{tab:DQ}
\end{deluxetable}

Second, hydrodynamical models predict that circumbinary disk material will
periodically form an accretion stream that crosses the cleared gap to feed
circumstellar disks or accrete directly onto the stars themselves
\citep{Artymowicz&Lubow1996}. Observations of ongoing accretion in pre-MS
binary stars necessitates this refueling behavior to balance the short
timescale on which a dynamically truncated circumstellar disk would be
exhausted through viscous accretion.

Driven by binary orbital motion, predictions for the frequency of circumbinary
accretion streams and their impact on stellar accretion rates are highly
dependent on the binary orbital parameters
\citep{Gunther&Kley2002,DeVal-Borroetal2011,GomezdeCastroetal2013}. Orbital
eccentricity in particular has a large effect where, for a given mass ratio,
the amplitude and ``sharpness'' of accretion events (in orbital phase) are
predicted to increase with increasing eccentricity. \citet[hereafter
  ML2016]{Munoz&Lai2016} for instance predict that equal-mass, circular
binaries will experience long-duration (multiple orbital periods) accretion
enhancements that occur every $\sim$5 orbital periods with a factor of 2
increase in the accretion rate at peak. A highly eccentric equal-mass binary,
on the other hand, is predicted to exhibit sharp accretion events every orbit
that evolve over roughly one-third of the orbital period and increase the
accretion rate by more than a factor of 10 at peak. With these orbital
parameter dependencies, short-period, eccentric systems provide the best
opportunity to test accretion models.

Focusing on this advantageous corner of the eccentricity-period parameter
space (analogous to DQ Tau; Table \ref{tab:DQ}), the general consensus of
models is that each apastron passage (orbital phase $\phi$=0.5) will induce a
stream of material from the circumbinary disk (CBD) that feeds a burst of
accretion during periastron passage ($\phi$=0,1). The specific morphology and
amplitude of the accretion events varies from one modeling effort to the next
(i.e. saw-toothed vs. symmetric rise and decay). Also, binary accretion
simulations to date have yet to include a magnetohydrodynamic (MHD) treatment,
which undoubtedly plays an important role close to the stars
(e.g. \citealt{Kulkarni&Romanova2008}). If these models are representative of
binary accretion, they would imply very different angular momentum histories
compared to single stars and a more dynamic disk environment relevant for
planet formation.

\subsection{DQ Tau}
\label{DQTau}

Since its discovery as a pre-main sequence (pre-MS) spectroscopic binary, DQ
Tau has become one of the primary targets in confronting theory of the
binary-disk interaction \citep{Mathieuetal1997,Basrietal1997}. Meeting the
criteria of a classical T Tauri star (CTTS) with evidence of ongoing accretion
and a gaseous protoplanetary disk, DQ Tau is one of a few, well-characterized
pre-MS binary systems capable of informing the physics of star and planet
formation in the binary environment.

The most extensive characterization of DQ Tau comes from
\citet{Czekalaetal2016}. Their study combines the orbital solution from
high-resolution, optical spectroscopy with disk kinematics derived from ALMA
observations to jointly constrain the orbital parameters, stellar
characteristics, and critically, the orbital inclination of the system. We
compile their results and other relevant system parameters from other works in
Table \ref{tab:DQ}.

DQ Tau was the first source to provide observational evidence for the pulsed
accretion theory. At many, but not all, periastron passages the system
exhibited sharp increases in both broadband and H$\alpha$ luminosities
\citep{Mathieuetal1997,Basrietal1997}, the same orbital phase predicted by
simulations with DQ Tau's orbital parameters \citep{Artymowicz&Lubow1996}.
Broad and variable H$\alpha$ emission line profiles provided support that
accretion was, at least in part, the source of the photometric
variability. Subsequent studies in the NIR also supported the pulsed accretion
interpretation with detections of diffuse, warm gas within a cleared central
cavity \citep{Carretal2001,Bodenetal2009}. These results were limited however
in their temporal and/or wavelength coverage. Sparse spectroscopic and
interferometric observations provide valuable snap-shots of the system but are
unable to test the temporal predictions of binary accretion theory. Even the
\citet{Mathieuetal1997} $V$-band photometry ($\sim$10 observations per orbit)
was only marginally sensitive to accretion, compared to $U$-band for instance
(e.g. \citealt{Venutietal2014}), and lacked the time-resolution necessary to
test accretion models in detail.

While the above studies provide encouraging results for pulsed accretion
theory, the quasi-periodic broadband, photometric behavior observed in DQ Tau
is not exclusive to periodic enhanced accretion events alone. Magnetic
reconnection events on low mass stars can create optical flares with the same
general broadband characteristics of accretion. During magnetic reconnection,
magnetic energy is converted into kinetic energy accelerating electrons along
field lines. In stellar flares, these flows impact the chromosphere and
photosphere where relativistic electrons deposit their energy creating a
photospheric hot-spot and white-light excess very similar to that of accretion
(e.g. compare \citealt{Kowalskietal2013} and
\citealt{Herczeg&Hillenbrand2008}). Stellar flares are stochastic events but
in a highly eccentric binary like DQ Tau, orbital motion brings the stars from
$\sim$43 stellar radii ($R_\star$) at apastron to $\sim$12$R_\star$ at closest
approach where the collision between each star's magnetosphere may induce a
series of magnetic reconnection events. \citet{Salteretal2010} find evidence
for such events with observations of recurrent synchrotron, mm-wave flares
(typical of stellar/solar flares) near the periastron passages of DQ Tau. If
these events are capable of depositing their energy in the stellar surface, a
large magnetic reconnection event or series of them could create optical
flares near periastron that masquerade as the signal of periodic enhanced
accretion in low-cadence broadband photometry. High-cadence, multi-color
photometry, however, can distinguish between stellar flares and accretion
variability.

In an effort determine the primary physical mechanism behind DQ Tau's
photometric variability, we have carried out an extensive monitoring campaign
combining moderate and high-cadence optical photometry spanning more than 10
orbital periods. Our observations are capable of detecting and characterizing
periodic pulsed accretion while determining the contribution from magnetic
reconnection events. By monitoring the accretion rate as a function of orbital
phase, these data provide a direct test of binary accretion theory and will
extend our understanding of the star-disk interaction to binary systems.

A description of our observations is provided in Section \ref{obsdr} as well
as our data reduction and calibration procedures. In Section \ref{detmec} we
discuss the morphology of our lightcurves and determine the dominant physical
mechanism behind DQ Tau's variability. We also characterize magnetic
reconnection events and their frequency, and place our results in context of
the colliding magnetosphere scenario. In Section \ref{Accretion} we calculate
mass accretion rates, establish the presence of periodic enhanced accretion
events, and comment on their variability. Section \ref{conc} provides a
summary of our results.

\section{OBSERVATIONS \& DATA REDUCTION}
\label{obsdr}

Observations capable of detecting and characterizing pulsed accretion events
in pre-MS binaries require multi-color photometric coverage over many orbital
cycles at a cadence that is a fraction of the orbital period. These formidable
demands are well met by the capabilities of the Las Cumbres Observatories
Global Telescope (LCOGT) Network \citep{Brownetal2013}. Described below
(Section \ref{lcogt}), these data form the basis of our observational study of
DQ Tau.

Despite the comprehensive nature of our LCOGT observations, they are not
capable of characterizing short-timescale events such as flares. To gain
sensitivity in this time domain, we supplement our moderate-cadence LCOGT
observations with 33 nights of concurrent minute-cadence, multi-color
photometry centered on 4 separate periastron passages. These single-site,
traditional observing runs were carried out at the WIYN 0.9m\footnote{The WIYN
  Observatory is a joint facility of the University of Wisconsin-Madison,
  Indiana University, the National Optical Astronomy Observatory and the
  University of Missouri.} (Section \ref{wiyn}) and ARCSAT 0.5m (Section
\ref{arcsat}) telescopes. At the end of this section, we describe our
photometry (Section \ref{photometry}) and calibration (Section \ref{stand})
schemes.

\subsection{LCOGT 1m Network}
\label{lcogt}

The LCOGT 1m network consists of 9 1m telescopes spread across 4 international
sites: McDonald Observatory (USA), CTIO (Chile), SAAO (South Africa), and
Siding Springs Observatory (Australia). Together, they provide near-continuous
coverage of the southern sky with automated queue-scheduled observing. At the
time of our observations, a majority of the 1m network was outfitted with
identical SBIG imagers which were chosen to maximize observing
efficiency. These 4k$\times$4k CCD imagers have 15.8$\arcmin$ fields-of-view
with 0.464$\arcsec$ pixels in standard 2$\times$2 binning.

Over the 2014-2015 winter observing season, our program requested queued
``visits'' of DQ Tau 20 times per orbital cycle for 10 continuous orbital
periods. Given the orbital period of DQ Tau, the visit cadence corresponded to
$\sim$20 hours. Each visit consisted of 3 observations in each of the
broadband $UBVRIY$ and narrow-band H$\alpha$ and H$\beta$ filters requiring
$\sim$20 min. The execution of our program went exceeding well with 218
completed visits made over 163 days ($\sim$10.3 orbital periods) with a mean
time between visits of 18.0 hours.

Observations are automatically reduced by the LCOGT
pipeline\footnote{https://lcogt.net/observatory/data/pipeline/}, which
performs bad-pixel masking, bias and dark subtraction, and flat-field
correction. The three images per filter are then aligned, median combined, and
fit with astrometric solutions using standard IRAF\footnote{IRAF is
  distributed by the National Optical Astronomy Observatory, which is operated
  by the Association of Universities for Research in Astronomy (AURA) under a
  cooperative agreement with the National Science Foundation.} tasks.

While observations were made in all of the filters listed above, in this work
we present only those in $UBVR$, which overlap with our high-cadence
observations described below. The full observational data set for DQ Tau and
other pre-MS binaries in our LCOGT observing campaign will be presented in an
upcoming paper.

Figure \ref{fig:lc} presents our LCOGT, $UBVR$ lightcurves in $\times$ symbols
plotted against an arbitrary orbital cycle number beginning at the start of
our observations.

\subsection{WIYN 0.9m}
\label{wiyn}

Two eight-night observing runs centered on separate periastron passages of DQ
Tau (orbital cycles 3 and 5 in Figure \ref{fig:lc}) were obtained from the
WIYN 0.9m telescope located at the Kitt Peak National
Observatory. Observations were made cycling through the $UBVR$ filters to
achieve the highest cadence possible while maintaining a signal-to-noise ratio
of $\sim$100 per stellar point-spread-function.

Our first run obtained some amount of data on all 8 nights. The first 6 of
these nights used the S2KB imager while the standard Half-Degree
Imager\footnote{http://www.noao.edu/0.9m/observe/hdi/hdi{\_}manual.html} (HDI)
was being serviced. S2KB is a 2048$^2$ CCD with a 20.48$\arcmin$ field-of-view
(FOV) and 0.6$\arcsec$ pixels. Binning (2$\times$2) and chip windowing
($\sim$10$\arcmin$) were implemented to reduce the readout time and increase
our observing cadence. With these measures the average filter cycle cadence
was reduced to $\sim$5.5 minutes.

HDI was used for the remaining two nights of our first run. HDI is a
4k$\times$4k CCD with a 29.2$\arcmin$ FOV and 0.43$\arcsec$ pixels. Using the
four-amplifier mode we were able to reach an improved observing cadence of
$\sim$3.6 minutes per filter cycle. Our second run utilized HDI exclusively
and obtained observations on 6 of the 8 nights. Data from both observing runs
were bias subtracted, flat-field corrected, and fit with astrometric solutions
using standard IRAF tasks.

In addition to our two eight-night observing runs, a synoptic observation
program was also in place at the WIYN 0.9m that provided $\sim$weekly
observations of DQ Tau in $UBVR$ during the 2014-B semester.

\subsection{ARCSAT 0.5m}
\label{arcsat}

Using Apache Point Observatory's ARCSAT 0.5m telescope, we performed a 7 and
10 night observing run centered on two separate periastron passaged of DQ Tau
(orbital cycles 2 and 7 in Figure \ref{fig:lc}).  The 1024$\times$1024
FlareCam\footnote{http://www.apo.nmsu.edu/Telescopes/ARCSAT/\\ Instruments/arcsat{\_}instruments.html}
imager was used for both observing runs (11.2$\arcmin$ FOV; 0.66$\arcsec$
pixels). Cycling through the SDSS $u^\prime g^\prime r^\prime i^\prime$
filters (Johnson filters were not available) provided an average cadence of
$\sim$3.8 minutes per filter cycle.

Our first observing run obtained observations on 5 of the 7 nights and 8 of
the 10 nights on the second. Data from these runs were bias and dark
subtracted, flat field corrected, and fit with an astrometric solution using
standard IRAF tasks.

\subsection{Photometry}
\label{photometry}

Given the large number of images obtained for this project, we rely on the
SExtractor \citep{Bertin&Arnouts1996} software to perform automated source
detection and aperture photometry. For each individual data set (LCOGT,
ARCSAT, WIYN 0.9m HDI, WIYN 0.9m S2KB) a matched catalog of each star's
instrumental magnitude is created image-by-image. This catalog is used to
perform ensemble photometry following the \citet{Honeycutt1992} formalism in a
custom Python implementation.

In short, a system of linear equations is solved to minimize the variation of
all stars within our catalog, weighted by their signal-to-noise. Variable, or
non-constant stars (including the target) are then interactively removed from
the system of equations based on their standard deviation compared to stars of
similar magnitude. Iteratively, stars are removed from the solution until only
steady, non-varying comparison stars remain, producing differential-lightcurve
magnitudes for all stars. We require a minimum of 3, non-variable comparison
stars for each image, and each comparison star must be present in at least 30
separate images across the data set. This technique is ideal for our highly
inhomogeneous observations in which observing conditions or pointing errors
may change the number and/or collection of comparisons stars available in a
given image.

\subsection{Photometric Calibration}
\label{stand}

Once differential magnitudes are derived for each individual dataset, we
perform the photometric calibration required to make direct comparisons across
datasets and to calculate mass accretion rates from a measure of the accretion
luminosity. While we did not observe traditional standard stars during our
observing program, the large FOV of HDI includes three stars for which
\citet{Pickles&Depagne2010} have produced ``fitted'' apparent magnitudes. By
fitting the published Tycho2 $B_TV_T$, NOMAD $R_N$, and 2MASS $JHK$ data with
a library of observed, flux-calibrated spectra, these authors have produced
best-fit apparent broadband photometry for 2.4 million stars. The 1$\sigma$
errors on each star's best-fit magnitudes are $\sim$0.2, 0.06, 0.04, and 0.04
mag for $UBVR$, respectively. The three stars used in our calibration have the
following Tycho2 IDs: TYC 1271-1341-1, TYC 1284-216-1, TYC 1271-1195-1. Their
best-fit-magnitudes range from 9.76 to 11.65 in $V$-band magnitude and 0.76 to
1.67 in $B-V$ color.

Using these three stars as our standard calibrators we calculate magnitude
zero-points and color coefficients during a photometric night of our HDI
run. RMS values from color-magnitude relations were on par or less than the
errors quoted in \citet{Pickles&Depagne2010}. They are 0.24, 0.10, 0.05 and
0.07 mag for $UBVR$, respectively. As these stars are only observed in the HDI
FOV, we use them to measure apparent magnitudes for all non-variable
comparison stars near DQ Tau which are then used to standardize the smaller
FOVs of the LCOGT and S2KB datasets.

In the case of ARCSAT, only the SDSS $u^\prime g^\prime r^\prime i^\prime$
filters were available for our observations. To convert these data to the
Johnson filter system, the \citet{Jesteretal2005} Johnson-to-SDSS
transformations were used to place our newly calibrated comparison stars into
the Sloan system. These were then used to calibrated the differential Sloan
magnitudes from ensemble photometry before finally transforming them into the
Johnson system.

\begin{figure*}[!tbh]
  \centering \includegraphics[keepaspectratio=true,scale=1.0]{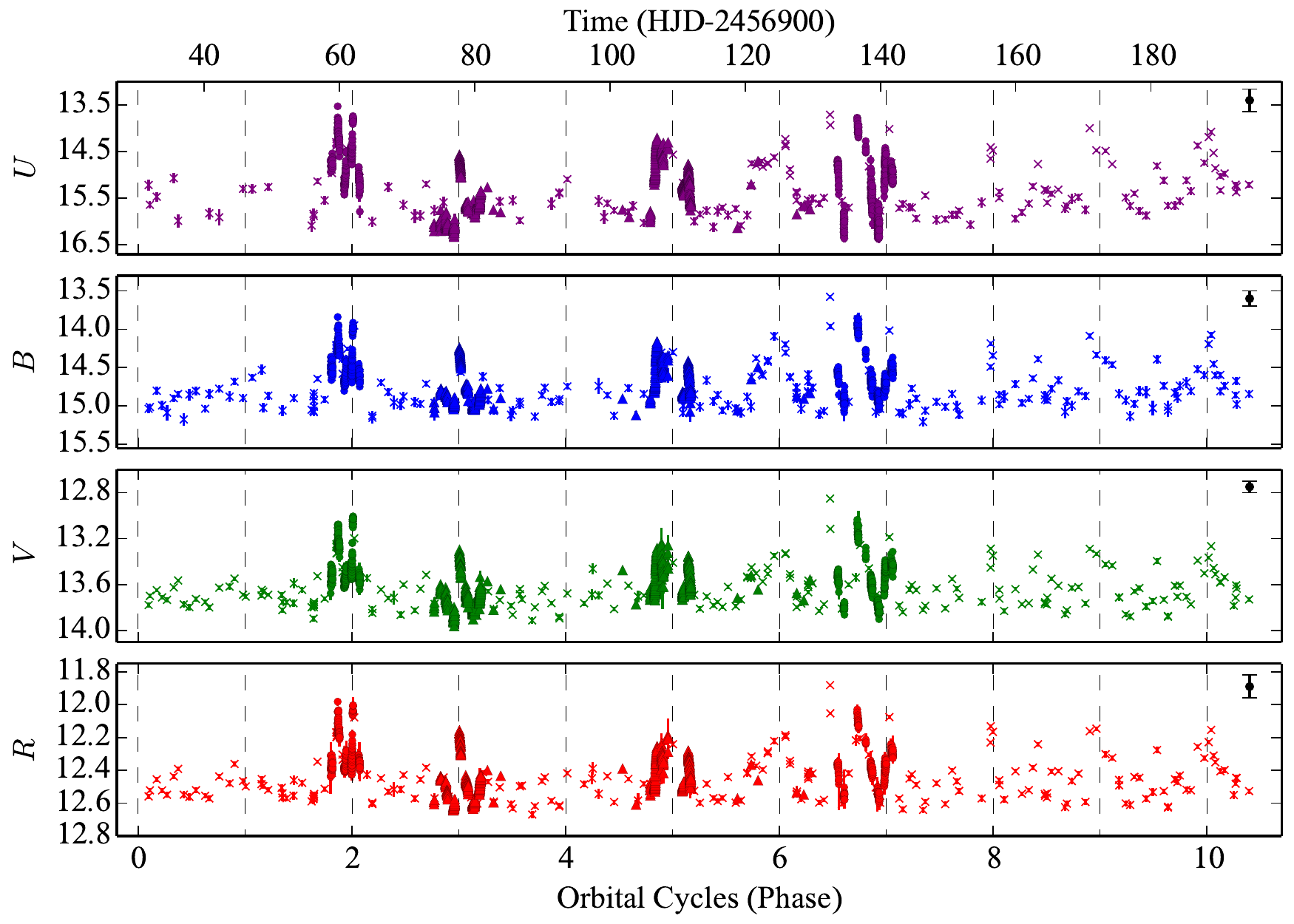}
    \caption{DQ Tau $UBVR$ lightcurves plotted against (arbitrary) orbital
      cycle number. LCOGT, ARCSAT, and WIYN 0.9m data are represented as
      $\times$, circle, and triangle symbols, respectively. ARCSAT data are
      transformed from SDSS to Johnson filters. Vertical dashed lines mark
      periastron passages. Note difference in the y-axis scale for different
      filters. Large error bar in the top right of each panel represents the
      systematic error in our photometric calibration. (A machine-readable
      table of the data presented in this figure is available in the online
      journal.)}
    \label{fig:lc}
\end{figure*}

Near-simultaneous LCOGT and high-cadence observations provide the opportunity
to directly test the agreement of our calibration between datasets. Comparing
observations made within 20 minutes of each other (typically 6), the LCOGT-HDI
and LCOGT-S2KB mean offsets agree to less than the uncertainties quoted in
\citet{Pickles&Depagne2010} for each filter. The LCOGT-ARCSAT offsets are
larger, owing to the additional transformation, but are still modest; 0.10,
0.20, 0.10, 0.02 mag for the $UBVR$ transformed magnitudes, respectively. A
final offset was applied to match zero-point variations to the HDI dataset
from which the apparent magnitudes are initially derived. Offsets were first
calculated for overlapping HDI-LCOGT data and then extended to the WIYN 0.9m
and ARCSAT datasets (overlapping with LCOGT).

The systematic errors involved in our calibration procedure are much larger
than the random error on any given point and the random errors are small
compared to the intrinsic variability observed. To remain cognizant of the
systematic errors however, we propagate them through each step of our analysis
and present them as the black error bar in the top right corner of Figures
\ref{fig:lc}, \ref{fig:accretion}, \ref{fig:lcPAA}, and \ref{fig:M_dot_vary}.

A machine-readable table providing the epoch of observation (Heliocentric
Julian date), zero-point corrected apparent magnitude, random magnitude error,
and observing facility for each of the $UBVR$ filters can be found in the
online journal associated with Figure \ref{fig:lc}.

\section{DETERMINING OPTICAL EMISSION MECHANISMS}
\label{detmec}

The optical emission from accretion and from stellar flares is dominated by a
combination of Balmer continuum emission and blackbody radiation. During
accretion, the flow of disk material along magnetic flux tubes approaches
free-fall velocity (supersonic) toward the stellar surface creating a standing
shock above the photosphere. Optically thin material in the post-shock region
is responsible for a majority of the blue-optical emission in the form of
Balmer continuum. Beneath the post-shock region, the photosphere is
radiatively heated creating excess blackbody emission from a hot-spot
\citep{Calvet&Gullbring1998}. Hot-spot temperatures have been modeled ranging
from 6500 to 10500 K for late M spectral type stars ($\sim$3000 K photospheric
temperatures) with most temperatures in the 8000 to 9000 K range
\citep{Herczeg&Hillenbrand2008}.

During a stellar flare, mass-loaded magnetic field lines in the chromosphere
or corona develop unstable configurations, leading to magnetic reconnection
events that accelerate charged particles towards the footpoints of the new
magnetic configuration. In the thick-target electron beam model used to
describe solar and stellar flares \citep{Brown1971}, these relativistic
particles interact with the chromosphere and photosphere where they deposit
their energy creating a white-light excesses \citep{Allredetal2006}. While the
mechanism by which mechanical energy is converted into radiative energy
remains an open question, most solar/stellar flares follow this general model
\citep[and references therein]{Fletcheretal2011}. Observationally, the
blackbody component of stellar flares dominates over Balmer continuum at the
flare peak where hot-spot temperatures range between 10000 and 14000 K,
reducing to 7000 to 10000 K in the decay phase \citep{Kowalskietal2013}. The
higher blackbody temperatures compared to accretion result from energy
deposition directly into the photosphere by the electron beam rather than from
radiative heating \citep{Kowalskietal2015}.

While both accretion and flares emit optical light by depositing energy and
mass into the stellar surface, the timescale, morphology, and detailed SED of
each processes variation can be distinguished with high-cadence, multi-color
optical photometry. Accretion is observed to naturally occur in bursts above a
steady accretion rate lasting days at a time without a consistent lightcurve
morphology \citep{Staufferetal2014}. This timescale may be related to the time
for instabilities to develop at the disk-magnetosphere interface
\citep{Kulkarni&Romanova2008,Inglebyetal2015}.

Stellar flare morphologies on the other hand have been extensively
characterized in the case of active M dwarfs, through high-cadence,
uninterrupted observation with the {\it Kepler} space telescope. Most flares
($\sim$ 85\%) exhibit the ``classical'' morphology consisting of an
impulsive-rise followed by an exponential-decay (\citealt{Davenportetal2014};
their Figure 3). The ratio of rise-to-decay times varies from $\sim$0.05 to 1,
with rise times typically shorter than 10 minutes. Flares also come in
non-classical flavors: ``complex'' or ``hybrid'', a superposition of multiple
classical flare events, and ``gradual'' or ``slow'', which are less impulsive
\citep{Kowalskietal2013,Dal&Evren2010}.  Regardless of the flare type, the
rise-times are generally less than 1 hour. For reference, the longest optical
flare observed on any star (M dwarf, pre-MS, or RS CVn) occurred over $\sim$10
hours and took $\sim$30 minutes to rise \citep[YZ CMi]{Kowalskietal2010}.

We focus on M dwarf flares because the combination of being intrinsically
faint (making it possible to detect small flares) and ubiquitous in the galaxy
has made them the subject of the most extensive flare studies to date. The
observed temporal and morphological characteristics, however, are consistent
with the more limited studies of stellar flares on pre-MS stars
\citep{Fernandezetal2004}, making them suitable for comparison with DQ
Tau. Pre-MS stars appear only to differ in that they have typical flare
energies that are a factor of 100 (or more) larger than M dwarfs.
 
The difference in color between accretion and stellar flares is more subtle
than that of the timescale and morphology, especially when only considering
the coarse wavelength information presented here ($UBVR$). In general, the
peak emission from a stellar flare is bluer than accretion radiation due to
the strong, high-temperature blackbody component. As the flare decays,
however, this distinction in color becomes less apparent.

To access the physical mechanism behind the broadband variability seen in DQ
Tau, we investigate the morphology, timescale, color, and energy associated
with brightening events. First, the qualitative aspects of the lightcurve
morphology and timescale of variation are compared to long-term, space-based
campaigns monitoring accreting young stellar objects and active M
dwarfs. Before characterizing the properties of accretion in Section
\ref{Accretion}, we define quantitative limits for the detection of flares,
characterize the color, timescales, and energy of those that are detected, and
place limits on their contribution to the total optical variability. Finally,
we place our results in the context of the colliding magnetospheres scenario.

\begin{figure*}[!tbh]
  \centering
    \includegraphics[keepaspectratio=true,scale=1.0]{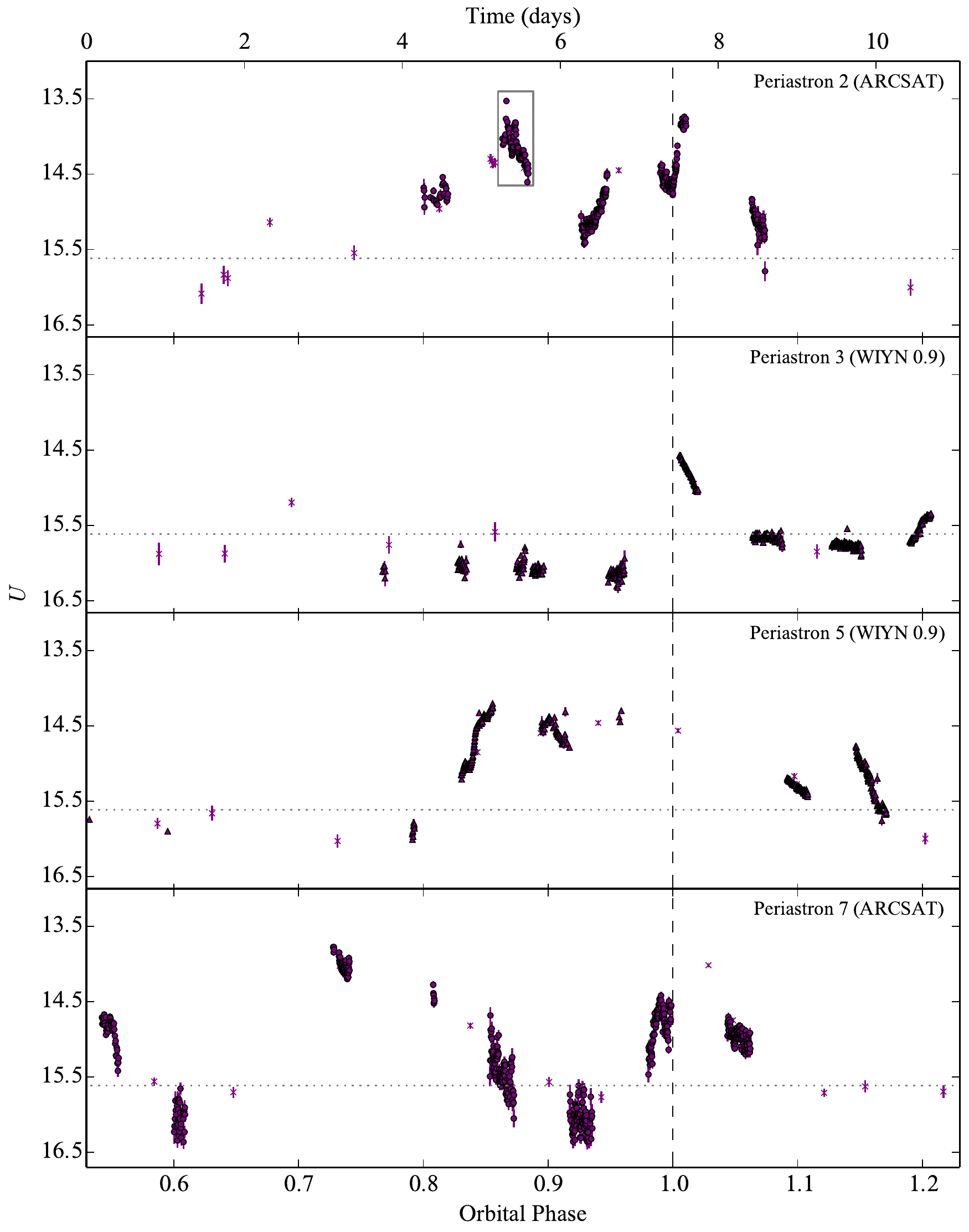}
    \caption{DQ Tau high-cadence, U-band lightcurves highlighting the rise and
      decline over the course of days near periastron passage. Vertical dashed
      lines mark periastron passage. Horizontal dotted lines mark the
      quiescent brightness. Periastron number in relation to Figure
      \ref{fig:lc} is provided in the top right of each panel along with the
      source of the data. LCOGT observations are $\times$ symbols. The light
      gray box in the top panel marks the region plotted in Figures
      \ref{fig:lczoom} and \ref{fig:lcflare}.}
    \label{fig:lchc}
\end{figure*}

\subsection{Lightcurve Characteristics}
\label{morph}

Figure \ref{fig:lc} presents our full DQ Tau, $UBVR$ lightcurves covering 10.3
orbital periods ($\sim$163 days). LCOGT observations are presented as $\times$
symbols with ARCSAT and WIYN 0.9m observations shown with circles and
triangles, respectively. The bottom x-axis is an arbitrary orbital cycle
number chosen to set the first observed periastron passage to 1. Below, we
refer to brightening events using the cycle number as it is presented in this
figure. (The top axis displays Heliocentric Julian days.)  Each periastron
passage is shown with a vertical dashed line. The y-axis scale of each panel
is set to match the variability of each filter and differs greatly with
photometric band. As expected in either accretion or flare events, the bluest
filters display the largest variability; $>$3 mag in $U$ while $<$1 mag in
$R$.

Focusing on the $U$-band lightcurve in Figure \ref{fig:lc} (our most sensitive
diagnostic of photospheric hot-spots, whether from accretion or flares),
brightening events of varying complexity and amplitude are seen around each
periastron passage. The duration of these events varies and can be as long as
half the orbital period. A significant amount of variability is also seen
outside of periastron, especially near certain apastron passages (e.g. orbital
cycles 6.5, 8.5, and 9.5).

Comparing our $V$-band lightcurve with that of \citet{Mathieuetal1997}, we
find consistent results with brightening events occurring around many, but not
all, periastron passages. Simultaneous observations in $U$-band, however,
reveal that ``quiescent'' $V$-band periastron passages do indeed have a
detectable $U$-band enhancement, due to the smaller contribution from the
stellar photospheres and a larger contribution from accretion luminosity at
shorter wavelengths. With the large range in time presented in Figure
\ref{fig:lc}, the detailed structure of brightening events are hard to discern
but already it is clear that some periastron passages display short, bursty
events (orbital cycles 3 and 8), while others display a prolonged elevated
state (orbital cycles 5, 6 and 9).

Figure \ref{fig:lchc} provides an expanded view of our high-cadence $U$-band
observations. Each panel presents a different periastron passage listed in the
top right, with vertical lines denoting the time of closest
approach. Horizontal dotted lines mark the quiescent $U$-band value from
orbital phases $\phi$=0.2 to 0.4 (consistently the quietest phase of the
orbit, see Figure \ref{fig:lcLSP} for reference). These data highlight the
complex structure of periastron brightening events showing variability in the
morphology, scale, and onset of the event. While variability is seen on a
variety of timescales, the underlying large scale evolution takes place over
days rather than hours. Each periastron passage observed with high-cadence
photometry shows increases above the quiescent level for tens of hours if not
days at a time.

Recent space-based campaigns monitoring the variability of accreting young
stellar objects and magnetically active M dwarfs provide a wealth of data
against which to compare our high-cadence observations. The {\it CoRoT} space
telescope monitored the star-forming region NGC 2264 for $\sim$40 days
continuously at a 512-second cadence, revealing a myriad of complex
variability trends \citep{Codyetal2014}. Comparing our $R$-band observations
to the $CoRoT$ $R$-band (white-light) lightcurves, we find many similarities
with the class of objects defined as ``bursters'' (\citealt{Staufferetal2014};
their Figure 1, right panels). These objects make up the dominant lightcurve
class of stars with large UV-excesses and are interpreted as episodic bursts
of accretion evolving over days at the few tenths of a magnitude level in
$CoRoT$ $R$. The variable morphology of these events as well as their
amplitude and timescale, support an accretion dominated interpretation of the
observed optical variability.

We also compare our lightcurves to the \citet{Hawleyetal2014} study of active
M dwarfs using $Kepler$, minute-cadence data. Variability in these stars is
dominated by sinusoidal star-spot modulations with sharp enhancements from
flares. Flares of this type would appear as near-vertical brightening events
in Figures \ref{fig:lc} and \ref{fig:lchc} while the observed enhancements are
smoother in nature.

\begin{figure}[!t]
  \centering
    \includegraphics[keepaspectratio=true,scale=0.5]{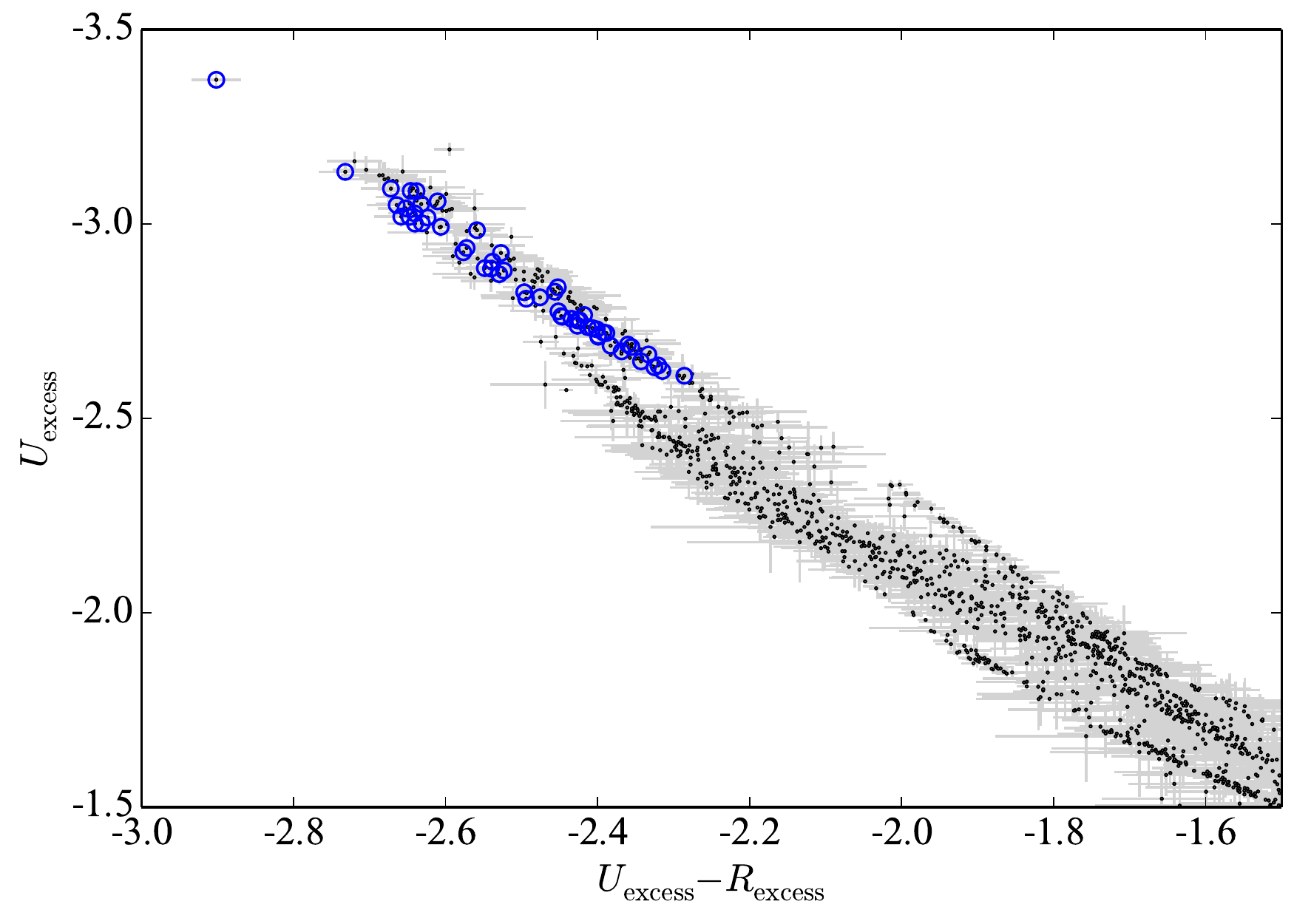}
    \caption{Extinction corrected $U$--$R$ color-magnitude diagram of emission
      above the stellar photosphere. Encircled points highlight candidate
      flares from high-cadence lightcurve analysis. The bluest point observed
      in $U$--$R$ is the peak emission of ``Flare 1'' presented in Figure
      \ref{fig:lcflare} (see text). (Extinction and template determination
      discussed in Section \ref{Accretion}.)}
    \label{fig:UR}
\end{figure}

The color of the variability also points to accretion. The observed $R$-band
increases are on the order of $\sim$0.5 mag with $U$-band excesses of $\sim$2
mag. This color is redder than what is typical of stellar flares at their
peak. Flares with peak $R$-band enhancements of 0.5 mag are rare and
accompanied by $U$-band components of $>$4 mag
\citep{Hawley&Pettersen1991,Davenportetal2014}. Figure \ref{fig:UR} presents
the extinction corrected $U$-$R$ excess color vs $U$-band excess above a
photospheric model (described in Section \ref{Accretion}). Most data do not
reach the extremely blue $U$$-$$R$ colors typical of large flare peaks
($U$$-$$R$ $\sim$ 3.5).

We explore the presence of flares in more detail in the following section but
in general, conclude that large scale changes in the accretion rate are the
most plausible source of optical variability based on the morphology,
timescale, amplitude, and color of the events.

\subsection{Stellar Flares}
\label{flares}

Although accretion processes appear to dominate the large-scale optical
variability on day timescales, we also investigate our nightly, high-cadence
lightcurves to determine the contribution from stellar flares. Based on the
empirical M dwarf flare behavior described above, we develop a flare finding
scheme aimed at detecting impulsive brightening events on timescales of tens
of minutes in our $U$-band, high-cadence lightcurves. Our detection scheme is
as follows: for each $U$-band observation the median value of data and its
error within the prior 60 minutes is computed (typically 12 to 20 points given
our average cadences on each telescope/detector combination). Points falling
10 times above the median error are then visually inspected as possible
flares. This conservative value is taken to compensate for the large
underlying variability from accretion. Our flare detection threshold is
adaptive in this case and can range from $\Delta U$=0.04 to 1.58 with a median
value of 0.32 mags. Using a shorter averaging window of 30 minutes recovers
the same results.

Following this procedure, three groups of points fall above our 10$\sigma$
threshold. The first two are short-timescale events we select as flare
candidates and discuss in detail below. The third comes from the steep rise
prior to periastron passage 5 (Figure \ref{fig:lchc}, third panel, orbital
phase $\sim$0.83). While a spectacular event in and of itself, rising more
than 1 mag in $U$ over the course of $>$9.5 hours, we do not classify it as a
flare given the relatively long timescale over which it is evolving.

Figure \ref{fig:UR} presents the $U$--$R$ color-magnitude diagram of emission
above the stellar photosphere. Data from the two candidate flares are
over-plotted with blue circles. The bluest point observed occurred during the
peak of the first candidate flare and is significantly bluer than other
measurements that are attributed to accretion. This aligns with our
expectation that the peak brightness of a stellar flare will be bluer than the
emission from accretion. Most of the rise and decay phase, however, are
indistinguishable in color space from the rest of the optical (accretion)
variability.

\begin{figure}[!t]
  \centering
    \includegraphics[keepaspectratio=true,scale=0.5]{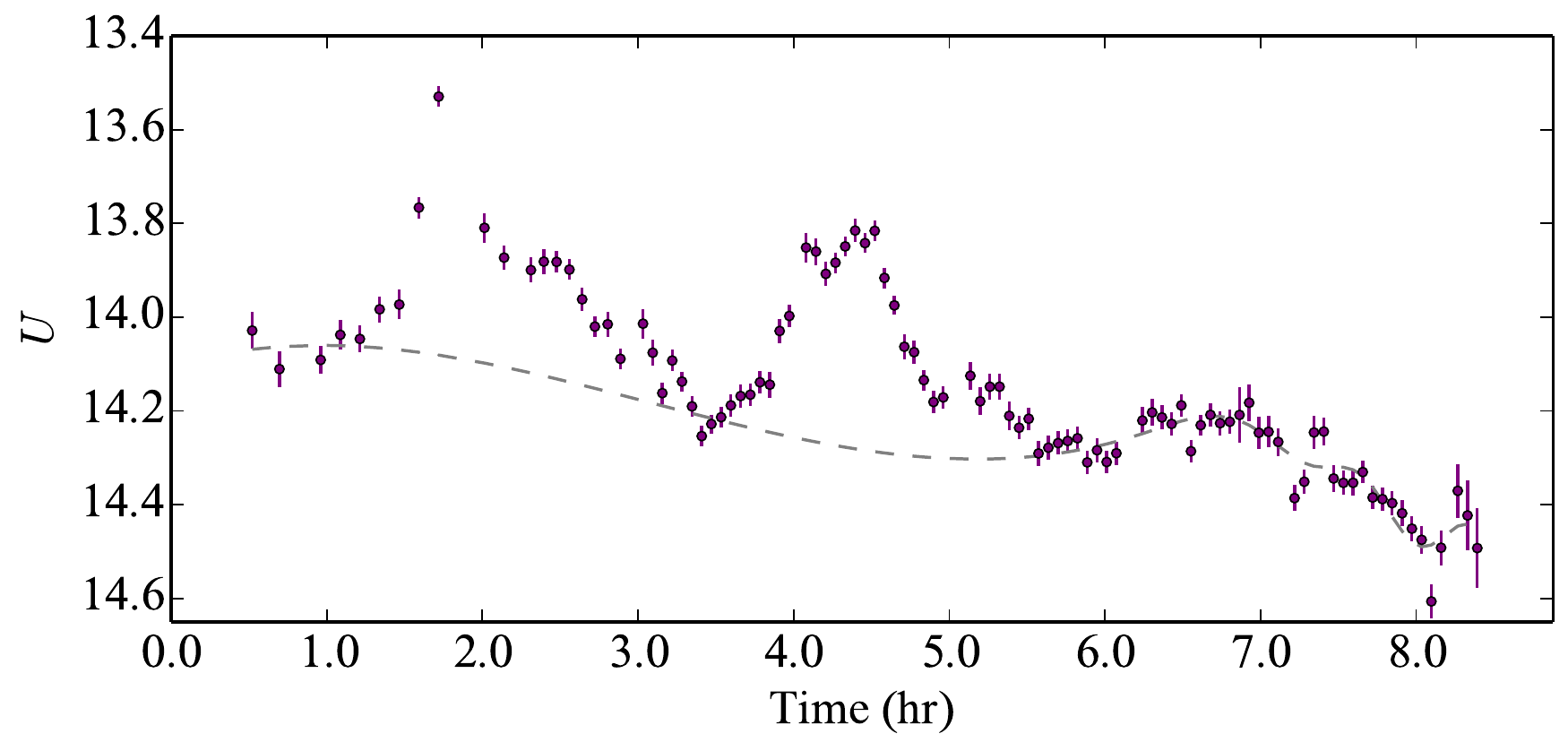}
    \caption{ARCSAT U-band lightcurve during one night of observation in which
      two flares are present. The gray box in the top panel of Figure
      \ref{fig:lchc} shows the location of these data with respect to orbital
      phase and the rest of the observing run. The gray dashed line displays a
      cubic spline fit to regions of the lightcurve devoid of flares.}
    \label{fig:lczoom}
\end{figure}

\begin{figure}[!t]
  \centering
    \includegraphics[keepaspectratio=true,scale=0.5]{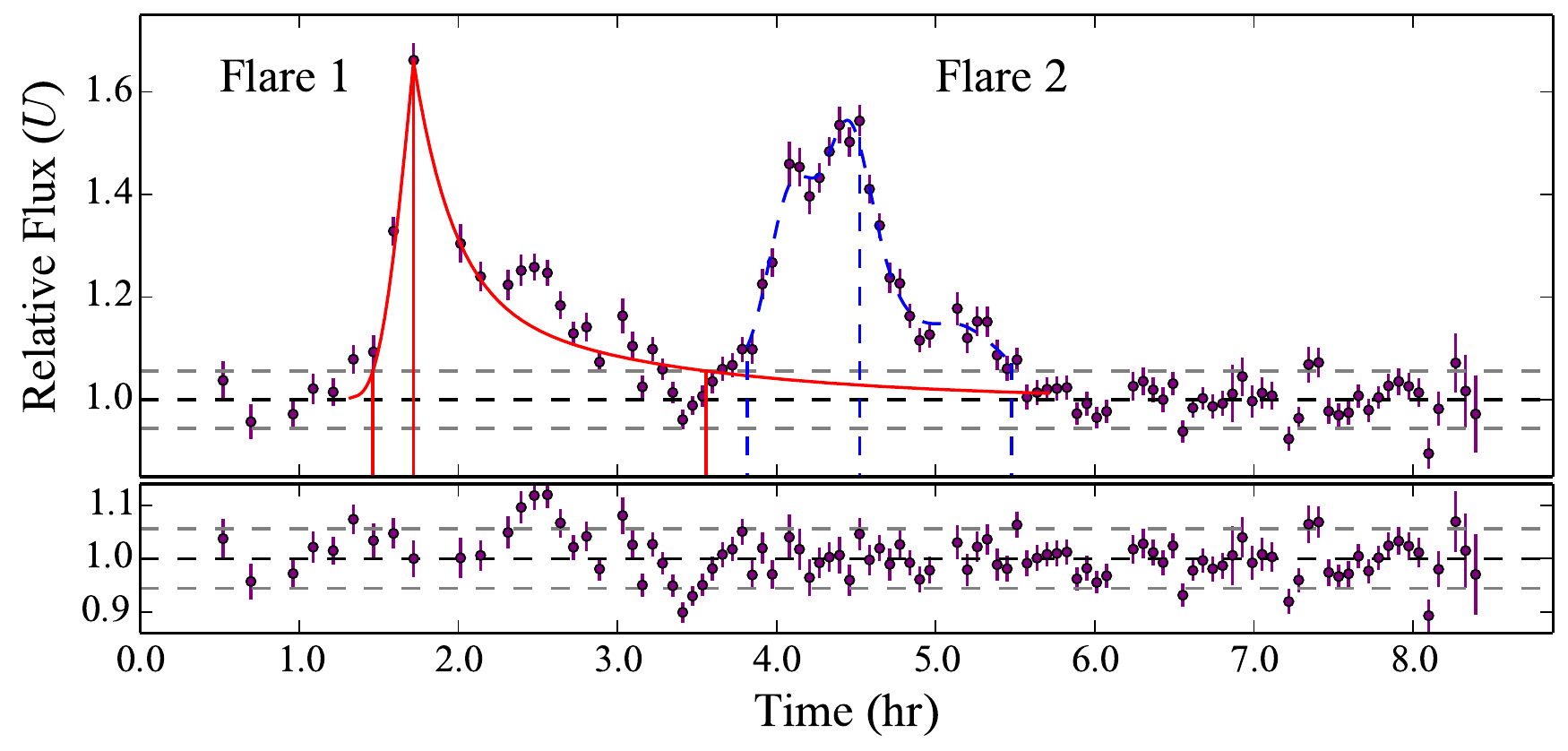}
    \caption{Lightcurve from Figure \ref{fig:lczoom} plotted as relative flux
      above the background accretion variability model. The red curve displays
      the classical flare template of \citet{Davenportetal2014}. Vertical red
      lines mark the beginning, peak, and end of the flare. The blue curve is
      a cubic spline fit to the classical flare subtracted lightcurve.
      Vertical blue dashed lines mark the beginning, peak, and end of the
      gradual flare. The bottom panel presents the residuals from model
      subtraction. Horizontal gray dashed lines in both panels mark the
      standard deviation of data devoid of flares after subtraction of the
      accretion model.}
    \label{fig:lcflare}
\end{figure}

Figure \ref{fig:lczoom} presents the night of high-cadence data in which our
candidate flares are detected. The fact that these two events fall close
together in time (partially overlapping) is not necessarily a concern given
that there is evidence for sympathetic flaring (flares triggering subsequent
flares) on low-mass stars \citep{Panagi&Andrews1995,Davenportetal2014} and the
Sun \citep{Pearce&Harrison1990}. To provide context within the large-scale
variability of DQ Tau, these data are highlighted in the top panel of Figure
\ref{fig:lchc} with a light gray box. In an attempt to characterize the
emission from these events alone, we fit a cubic spline to regions of the
lightcurve devoid of flares in order to remove accretion variability. The fit
is shown as the gray dashed line in Figure \ref{fig:lczoom}. Subtracting this
crude model and converting to normalized flux results in Figure
\ref{fig:lcflare}.

The first event in Figure \ref{fig:lcflare}, ``Flare 1'', has the morphology
of a classical flare. The red line over-plots an empirical classical flare
template from \citet{Davenportetal2014}. Constructed from 885 classical
white-light flares on the active M dwarf GJ 1243 observed with $Kepler$, this
flux normalized model is broken into a rise and decline phase that depends on
the event timescale, $t_{1/2}$, the time spent above half the peak flux. A
4th-order power series in $t_{1/2}$ describes the rise phase and a sum of two
exponentials describes the decline. We do not fit the template to our data in
a $\chi^2$ sense, but over-plot the template using the measured $t_{1/2}$
value and an amplitude normalization. The agreement is not perfect but given
the uncertainty in the background subtraction, we find it to be reasonable
evidence that this event is a flare from a magnetic reconnection event.

The second event during this night, ``Flare 2'', does not have the classical
flare morphology but may be a slow or gradual flare. Although, our cadence may
not be high enough to decompose multiple small classical flares events if it
were instead a hybrid or complex flare. Without an empirical model for
non-classical flares to compare against, we fit a cubic spline (blue curve in
Figure \ref{fig:lcflare}) to the accretion and classical flare subtracted
data.

In addition to the morphology and timescale arguments above, we make a
quantitative comparison of the flare energy in the $UBV$ filters to flares
observed on other pre-MS stars. We determine the rise and decay times for each
flare where our flare templates exceed the non-flaring standard deviation (top
dashed line in Figure \ref{fig:lcflare}; location of start, peak, and end
times are marked with vertical lines). The flare energy is then computed with
a trapezoidal integration of the excess emission above our accretion model
(dashed line is Figure \ref{fig:lcflare}) assuming a distance of 140pc and
$A_V$=1.5 (see Section \ref{Accretion}). Table \ref{tab:flares} presents their
temporal characteristics from the $U$-band lightcurve and total energy in the
$UBV$ filters. Error in the energy comes from applying the maximum and minimum
offsets of our photometric systematic error. The derived energies in each
filter fall within the spread of flares observed on other pre-MS stars
\citep[$2\times10^{34}<E_U$(ergs)$<1.1\times10^{36}$;][]{Gahm1990,Fernandezetal2004,Koen2015}
and the ratio of energy between each filter agrees with the trend seen on
pre-MS stars as well as M dwarfs \citep{Lacyetal1976,Gahm1990}. This result
provides further evidence for a magnetic reconnection origin of these events.

These two flares were the only events in our high-cadence lightcurves that had
the amplitude and timescale typical of magnetic reconnection as we understand
them from low-mass dwarfs and pre-MS stars. To quantitatively compare the
timescale of our flares to the large-scale variability, we measure the
$t_{1/2}$ values of the 10 largest brightening events observed at
high-cadence. Using the quiescent brightness level shown in Figure
\ref{fig:lchc} as the baseline, we find an average $t_{1/2}$ value of 21.7
hours with the shortest being 2.5 hours. These values are an order of
magnitude longer than those calculated for the flares in Table
\ref{tab:flares}.

\begin{deluxetable}{lcc}
\tablewidth{0pt}
\tabletypesize{\footnotesize}
\tablecaption{Flare Characterization Summary}
\tablehead{
  \colhead{Parameter} &
  \colhead{Flare 1} &
  \colhead{Flare 2}}
\startdata
$t_{1/2}$ (min)       & 23.9 & 41.6 \\
Rise Duration (min)  & 15.4 & 42.4 \\
Fall Duration (min)  & 110.2 & 57.4 \\
$\Delta U$ (mag)     & 0.55 & 0.47 \\
$U$-band Energy ($10^{35}$ ergs)  & $2.2\pm0.5$ & $1.6\pm0.4$\\
$B$-band Energy ($10^{35}$ ergs)  & $2.1\pm0.4$ & $1.0\pm0.2$\\
$V$-band Energy ($10^{35}$ ergs)  & $1.8\pm0.2$ & $0.5\pm0.1$\\
\enddata 
\tablecomments{Temporal measurements from $U$-band lightcurve.}
\label{tab:flares}
\end{deluxetable}

Lastly, to determine the fraction of our data in which flares are present, we
first calculate the amount of time in which our data are capable of detecting
flares. \citet{Hawleyetal2014} find that a majority of flares are less than
2.5 hours in duration. Setting this as the minimum duration of continuous
monitoring (with data gaps less than 30 minutes) required to detect flares,
141 hours of ``flare coverage'' are obtained. Within this window, only 4.1
hours contain flares at an average level of $\Delta U$$\sim$0.32,
corresponding to $\sim$3\%. Here we have assumed a perfect detection
efficiency above the detection threshold as each event is visually inspected
and characterized, finding it in good agreement with flares on other pre-MS
stars. With that in mind, this value should be taken as a lower limit on the
temporal flare contribution given our variable detection threshold. Small
flares that would go undetected in our data evolve quickly however, and would
not contribute significantly given the $\sim$3 magnitude $U$-band variations
observed in the system. We also note that the fraction of time spent flaring
derived above is from observation near periastron alone. Our data provide no
information on the occurrence of flares near apastron or if any
orbital-phase-dependence exists.

We conclude that flares play a very small roll in the amplitude and temporal
nature of DQ Tau's variability, and that the broadband variability is due to a
variable accretion rate. For the remainder of our discussion we remove the two
flares using the models described above (the residuals of which are shown in
the bottom panel of Figure \ref{fig:lcflare}) and attribute all remaining
variability to changes in the accretion rate.

\subsection{Colliding Magnetospheres}
\label{collide}

Here we consider whether the detection of flares near a periastron passage of
DQ Tau might be indicative of magnetic reconnection events in colliding
magnetospheres. In this scenario, the large-scale magnetic fields of both
stars interact during periastron approach (bringing the stars from $\sim$43 to
12 $R_\star$) leading to unstable magnetic configurations and reconnection in
the case of field lines with opposing polarity (see \citealt{Adamsetal2011}).

Evidence for colliding magnetospheric reconnection in DQ Tau comes from
\citet{Salteretal2010} who find recurrent, mm-wave synchrotron enhancements
during three out of four observed periastron passages. With only 8 to 16 hours
of observation per periastron passage, the consistency of radio flares points
to inter-magnetospheric reconnection being a commonplace event near
periastron. The largest of these events reached a peak luminosity of
$\sim10^{28}$ ergs s$^{-1}$ at 2.7 mm (115 GHz; 1 GHz bandpass) and while it
was not observed through its return to quiescence, the event was modeled with
a $\sim$30 hr duration. Radio flares of this amplitude have been observed on
the weak-lined T Tauri star (WTTS) binary V773 Tau
\citep{Massietal2002,Massietal2006}, which were also attributed to colliding
magnetospheres. Both, however, are an order of magnitude more luminous than
largest radio events observed on active M dwarfs \citep{Ostenetal2005} or RS
CVn binaries \citep{Trigilioetal1993}. If optical events similar to stellar
flares accompanied these events at amplitudes that scale with the radio
component, our observations would easily detect them given the sensitivity to
impulsive brightening events derived above.

While we have assumed that magnetic reconnection between colliding
magnetospheres is capable of creating an optical, stellar-flare-like
counterpart, determining the detailed characteristics of an optical
counterpart to radio events of this scale is difficult. Some of the most
extensive simultaneous radio and optical monitoring has been on active M
dwarfs. During flares the optical component is seen to evolve on a much
shorter timescale than the radio counterpart
\citep{Ostenetal2005,Butleretal2015}. The prolonged radio decay is attributed
to magnetic mirroring near footpoints where field lines converge, increasing
the field strength, and reflecting synchrotron producing electrons
(e.g. \citealt{Aschwandenetal1998}).  This effect may have a large impact on
magnetic reconnection events far from the stellar surface. The efficiency of
magnetic mirroring depends on the ratio of the field strengths that a particle
experiences which, for DQ Tau, assuming a simple dipole, would correspond to
$\sim$245 from 6.3$R_\star$ (midpoint between stars at periastron) to the
stellar surface. In solar flares where the site of reconnection is in the
chromosphere or corona, this ratio is typically measured as 2, or less
\citep{Tomczak&Ciborski2007,Aschwandenetal1998}.

Moving the site of reconnection further from the surface of the stars also
raises concerns of synchrotron radiative losses and the potential for
collisional losses with intervening circumstellar material that prevents
accelerated electrons from reaching the chromosphere. If the energy from
magnetic reconnection remains confined or lost to other processes it will
prevent the conversion of mechanical energy to an optical counterpart at the
stellar surface. \citet{Salteretal2010} present some simultaneous optical
photometry during the decay phase of one of their radio flares which also
shows a general decaying behavior (their Figure 7). While the match between
the optical and radio morphology is compelling, this behavior is not seen in
standard solar/stellar flares.

Aside from lightcurve morphology, we also compare the energy of optical and
radio brightening events to the available magnetic energy budget. Assuming
quasi-static, anti-aligned dipole fields, \citet{Adamsetal2011} estimate the
magnetic interaction energy available for reconnection events as a function of
the stellar radius, the surface magnetic field strength, and
apastron-to-periastron separation (their Equations 13 and 14). The interaction
energy is derived from the difference between the lowest energy magnetic field
configurations at periastron and apastron. Energy in this model is provided by
the orbital motion which compresses the fields and is only a fraction of the
total magnetic energy stored in the fields.

Adopting a surface dipole field strength of 1.5 kG and the parameters listed
in Table \ref{tab:DQ}, DQ Tau has an available interaction energy of
$\sim10^{35}$ ergs (only $\sim$1\% of the combined magnetic energy beyond an
interaction distance of 6.3$R_\star$ for each star). Integrating a synchrotron
source function matching the observed 90 GHz flux density from 0 to 90 GHz for
a range of power-law electron energy distributions (1.1 to 2.9), we find
energies ranging from 0.4-6.7$\times10^{35}$ ergs, assuming a 6.55 hr
$e$-folding decay timescale \citep{Salteretal2008,Salteretal2010}. For
comparison, trapezoidal integration of our photosphere-subtracted,
flux-calibrated observations produces an average of $\sim$10$^{38}$ ergs
emitted in the combined $UBVR$ filters during periastron passage ($\phi=0.7$
to 1.3), a factor of 10$^3$ more than the available magnetic energy budget.

Based on the multi-day variability of optical brightening events, the excess
of optical energy released near periastron when compared to the colliding
magnetosphere energy budget, the paucity of classical optical flare events
(for lack of a better model), and the favorable conditions for magnetic
mirroring, we conclude that reconnection events from colliding magnetospheres
do not contribute significantly to the periodic luminosity increases in our
optical lightcurves. The optical flares that are present do have energies that
agree with the colliding magnetosphere energy budget but, they are also
typical of flares on single pre-MS stars, are less regular than radio events,
and occur at a relatively wide stellar separations ($\sim$24$R_\star$). These
flares may very well be the result of magnetic reconnection on the surface of
one of the two stars. Simultaneous optical and radio observation will be
required, however, to make a definitive statement on their origin.

\section{Characterizing Accretion}
\label{Accretion}

\begin{figure*}[!t]
  \centering
  \includegraphics[keepaspectratio=true,scale=1.0]{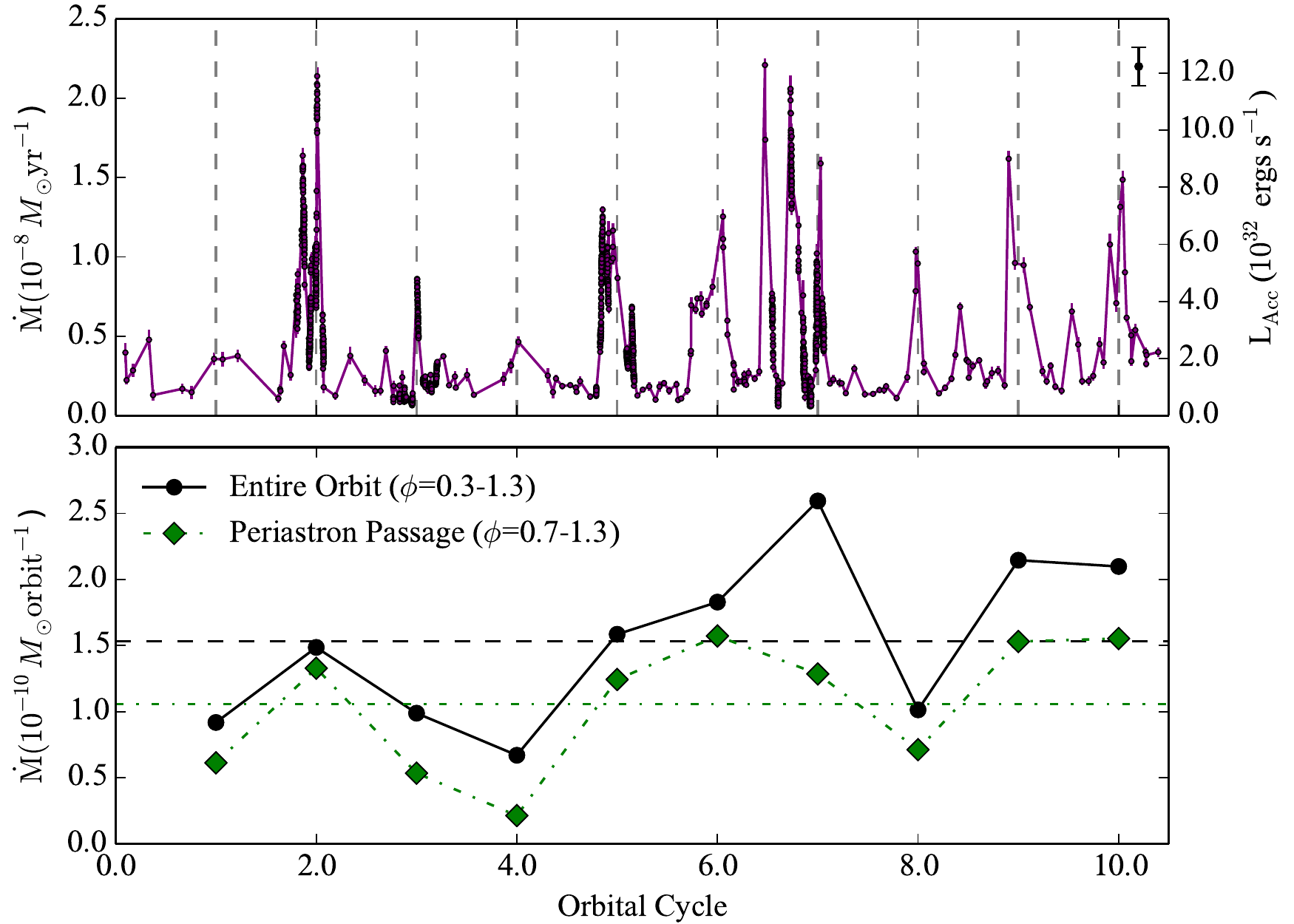}
  \caption{{\bf Top}: Accretion rate (left axis) as a function of orbital
    cycle. Right axis displays the corresponding accretion luminosity. Dashed
    vertical lines mark periastron passages. {\bf Bottom}: Integrated mass
    accreted per orbit. Black circles represent the mass accreted per full
    orbital period (orbital phases $\phi$=0.3 to 1.3). Green diamonds
    represent the mass accreted per periastron passage (orbital phases
    $\phi$=0.7 to 1.3). Horizontal dashed and dot-dashed lines mark the
    average orbital and periastron accreted masses, respectively.}
    \label{fig:accretion}
\end{figure*}

A measurement of the mass accretion rate can be made by determining the excess
emission above the stellar photosphere(s) resulting from accretion. This
requires an estimate of the underlying spectral type and extinction in the
absence of accretion. We determine these properties following the method
described in \citet{Herczeg&Hillenbrand2014}. These authors compute a library
of low-resolution, pre-MS spectral templates from a grid of 24 flux-calibrated
WTTS spectra, spanning spectral types K0 to M9.5. Empirical templates have the
advantage over synthetic spectra in that they include chromospheric emission
(see \citealt{Inglebyetal2011}) and provide more accurate colors for these
typically, highly spotted photospheres
\citep[e.g.][]{Grankinetal2008,Alencaretal2010}.  Templates are fit to the
spectra of accreting CTTSs modifying the intrinsic luminosity, extinction, and
additive accretion continuum level as free parameters.
\citet{Cardellietal1989} extinction curves are used assuming $R_V$=3.1 and the
accretion continuum is modeled as a constant flux value across wavelength. As
noted above, the true accretion spectrum has structure from the Balmer jump
and emission lines; however, the fits here only include wavelength regions
redward of 4000\AA \ and exclude emission lines. Within these
continuum-dominated windows, a flat spectrum provides an adequate description
of accretion while keeping the degrees of freedom minimal. The binary nature
of DQ Tau is ignored in this process but as a near equal-mass binary, the
combined spectrum of both stars should not differ greatly from that of a
single star at low-spectral resolution.

Applying this procedure to a flux-calibrated spectrum of DQ Tau obtained in
January 2008 with the Double Spectrograph \citep{Oke&Gunn1982} on the Hale 200
inch telescope (originally published in \citealt{Herczeg&Hillenbrand2014}), we
find a spectral type of M0.4 and an extinction of $A_V$=1.5. These values
agree with the results of \citet{Herczeg&Hillenbrand2014} who quote typical
uncertainties of 0.3 spectral type subclasses and 0.3 magnitudes of extinction
for (single) M stars. Both measurements also lie in the middle of the values
found in the literature
\citep{Strometal1989,Kenyon&Hartmann1995,Czekalaetal2016}.  Even though this
work is primarily concered with the relative changes of the accretion rate,
the importance of extinction on the derived accretion rate baseline should be
noted. The $\pm$0.3 magnitude uncertainty of this method corresponds to a 0.2
dex systematic uncertainty in all accretion luminosities (rates) and flare
luminosities (energies).

The WTTS templates extend from 3130-8707\AA \ with a central gap from
5689-6193\AA. Before convolving the best-fit template with filter curves we
fill this gap in the spectral coverage by finding the best-fit BT-Settl
atmospheric model \citep{Baraffeetal2015}. A best fit is found at a
temperature of 3900 K and log($g$) of 4.0, in agreement with
\citet{Czekalaetal2016}.

With a model for the combined photospheric contribution in DQ Tau, we
determine the mass accretion rate by first converting the $U$-band excess
luminosity into an accretion luminosity following the empirical relation found
by \citet{Gullbringetal1998}:
\begin{equation}
  {\rm log}(L_{\rm Acc}/L_\odot) = 
  1.09\ {\rm log}(L_{\rm {\it U_{\rm excess}}}/L_\odot)+0.98
\label{eqn:U2L}
\end{equation}
The $U$-band photospheric luminosity is computed by convolving the template
with a $U$-band filter curve \citep{MaizApellaniz2006,Pickles&Depagne2010},
adopting a distance of 140 pc. This luminosity is then subtracted from the
observed, extinction-corrected $U$-band luminosity, providing $L_{\rm {\it
    U_{\rm excess}}}$.

In these calculations we have ignored the contribution to variability from
star-spots. Spot variations on non-accreting pre-MS stars are typically a few
tenths of a magnitude in $U$-band \citep{Bouvieretal1995}. This is much
smaller than the observed variability and at a high inclination angle
($\sim$22 degrees), the geometry of hot and cool spot visibility due to
rotation should have a small effect.

From accretion luminosities we calculate mass accretion rates using the
following formula,
\begin{equation}
  \dot{M}\simeq \frac{L_{\rm Acc}R_\star}{GM_\star} \left(
  1-\frac{R_\star}{R_{\rm in}} \right)^{-1},
\label{eqn:MA}
\end{equation}
where $R_{\rm in}$ is the magnetospheric truncation radius from which
accreting material free-falls along field lines. The value of $R_{\rm in}$
depends on the strength of the magnetic field and the ram pressure of
accreting material. In the binary environment, where mass flows are predicted
to be highly variable and phase dependent (ML2016), the conditions of
accreting material are likely not well described by a single value of $R_{\rm
  in }$. As we discuss below, the ram pressure of accreting material is likely
highest near periastron. If this behavior corresponds to smaller $R_{\rm in}$
values, a constant value of $R_{\rm in}$ will underestimate the accretion rate
near periastron and overestimate it at times of low accreting ram pressure
(presumably apastron). Without a model for the time variable interaction of
the magnetic field with circumstellar material, we resort to the canonical
single star value of $R_{\rm in}=5R_\star$ \citep{Gullbringetal1998}, even
though it is less physically motivated in this case. Fortunately, the mass
accretion rate is fairly insensitive to $R_{\rm in}$ (a factor of 2 decrease
in $R_{\rm in}$ corresponds to a factor of 0.6 in the mass accretion
rate). Given these uncertainties, accretion luminosity measurements are also
included in Figures \ref{fig:accretion}, \ref{fig:lcLSP}, and
\ref{fig:M_dot_vary}.

Following this procedure we calculate mass accretion rates ranging from
$5.9\times10^{-10}$ to $2.2\times10^{-8} M_\odot {\rm yr}^{-1}$, in good
agreement with measurements from optical and NIR spectra
\citep{Gullbringetal1998,Bary&Petersen2014}. The top panel of Figure
\ref{fig:accretion} displays the mass accretion rate as a function of orbital
cycle. An increase in the accretion rate can been seen at every periastron
passage; at some, the accretion rate increases by more than a factor of 10
from the quiescent value.

The bottom panel of Figure \ref{fig:accretion} presents the mass accreted over
each full orbital period and over each periastron passage. For the full orbit,
we define our integration range to be orbital phases $\phi=0.3$ to 1.3 in
order to include the entire periastron event. For periastron passages, the
integration range is over orbital phase $\phi=0.7$ to 1.3. Black circles and
green diamonds mark the full orbit and periastron integrations, respectively,
with horizontal lines marking the mean of each. This periastron passage range
encloses 60\% of the orbital period but has a median contribution of 71\% to
the total mass accreted per orbital period. Large variability exists, however,
with periastron contributions ranging from 49-90\% of the total mass accreted
per orbital period.

\subsection{Periodic Enhanced Accretion}
\label{PEA}

\begin{figure}[!tb]
  \centering
    \includegraphics[keepaspectratio=true,scale=0.5]{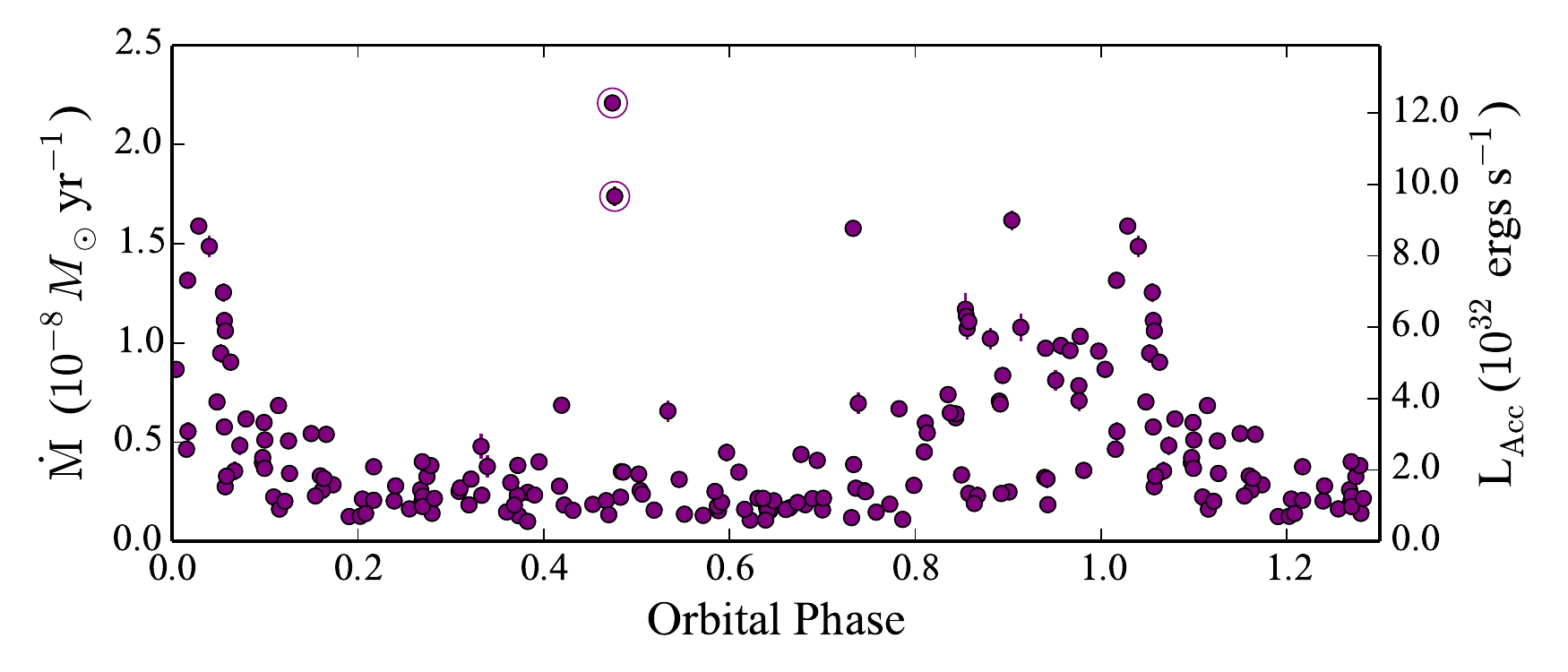}
    \includegraphics[keepaspectratio=true,scale=0.5]{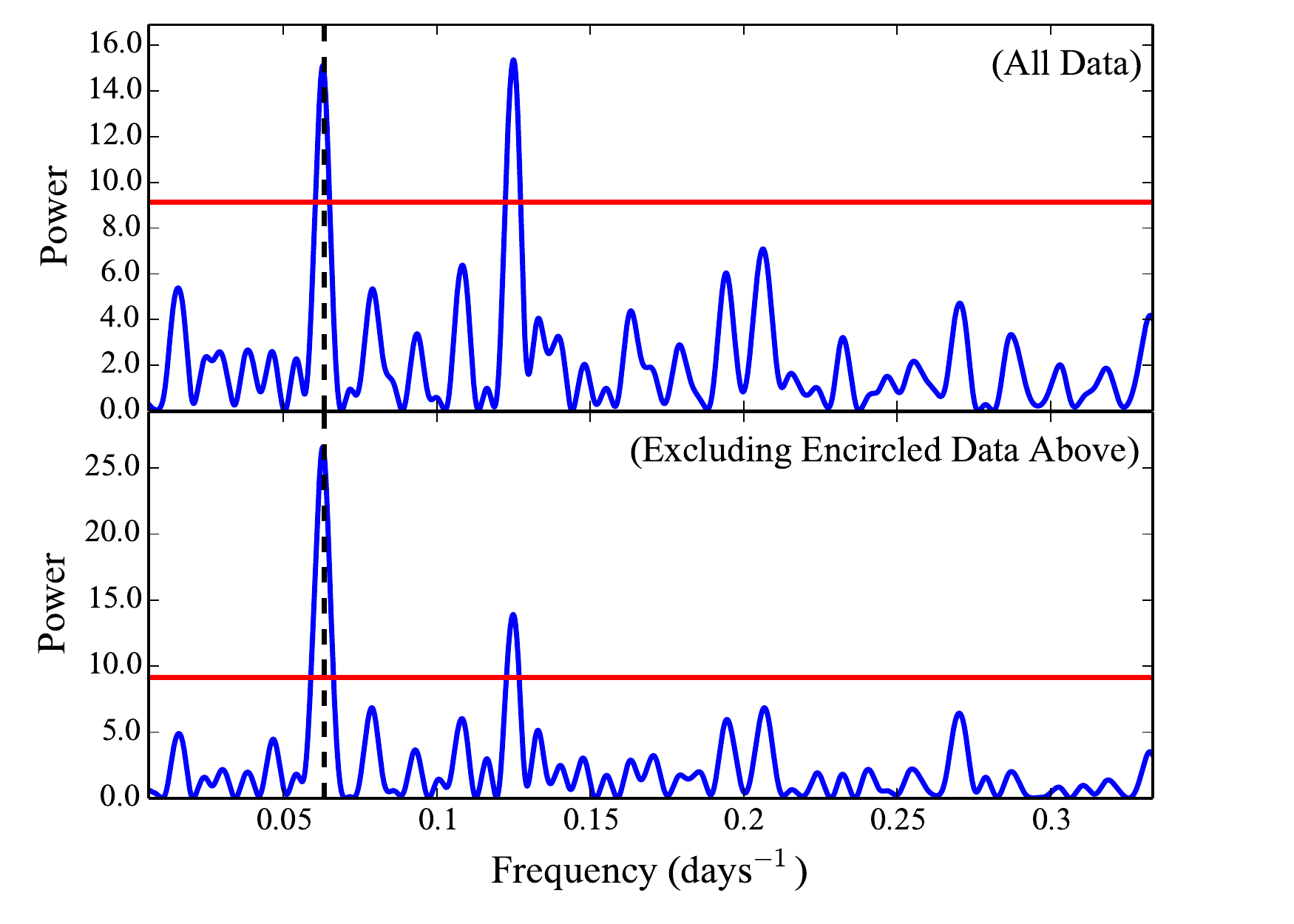}
    \caption{{\bf Top:} Mass accretion rate (left axis) from LCOGT
      observations phase-folded about the orbital period. Right axis displays
      the corresponding accretion luminosity {\bf Middle:} Lomb-Scargle
      periodogram of all of the above data. {\bf Bottom:} Lomb-Scargle
      periodogram of the above data excluding the two encircles periastron
      points. In both periodograms the horizontal red line marks the 99\%
      false-alarm-probability and the vertical dashed line is the
      radial-velocity determined orbital period.}
    \label{fig:lcLSP}
\end{figure}

Numerical simulations of the binary-disk interaction predict that in cases of
high eccentricity, discrete accretion events should occur near every
periastron passage. We test this prediction by performing a Lomb-Scargle
periodogram analysis \citep{Scargle1982} on the mass accretion rates derived
from LCOGT observations. Figure \ref{fig:lcLSP} displays the mass accretion
rate phase-folded about the spectroscopically determined orbital period in the
top panel and the periodogram of those data in the middle panel. The red line
in the bottom two panels marks the 99\% false-alarm-probability (FAP)
determined using a Monte Carlo bootstrap simulation \citep{Frescuraetal2008}.

Even with the large variability present near periastron, typical of accretion
in CTTSs, a significant peak is found near the spectroscopic period (marked
with the dashed vertical line). We find a period of 15.91$\pm$0.08 days, in
good agreement (1.3$\sigma$) with the orbital period. (Periodogram peak errors
are calculated by enclosing 68\% of a the probability-distribution-function
created from a $10^6$ iteration Monte Carlo, bootstrap simulation using
sampling with replacement in time and $\dot{M}$ \citep{Pressetal1992}.) This
spectral peak and the visual inspection of the Figure \ref{fig:accretion}
provide compelling evidence that, just as models predict, pulsed accretion
events occur periodically near each periastron passage.

A second significant peak found at half the orbital period is powered by
apastron accretion events. Most of the power at this frequency comes from the
two closely separated LCOGT observations near orbital cycle 6.5. These two
point are encircled in the top panel of Figure \ref{fig:lcLSP}. (Other
apastron accretion examples can be seen at orbital cycles 8.5 and 9.5 in
Figure \ref{fig:accretion}.) A periodogram excluding these two points is
presented in the bottom panel of Figure \ref{fig:lcLSP} where a peak is still
present above the 99\% FAP. Non-sinusoidal waveforms, like those observed, are
capable of producing harmonics above a 99\% FAP at integer multiples of the
primary frequency. This is potentially the case in the bottom panel of Figure
\ref{fig:lcLSP} but not in the middle panel where the peak at twice the
orbital frequency is the highest of the two. We conclude that apastron
accretion events are quasi-periodic, occurring at generally lower amplitudes
and with less consistency when compared to periastron accretion
events. Apastron accretion events are not predicted by the binary pulsed
accretion theory and are discussed further in Section \ref{apastacc}.

\begin{figure}[!t]
  \centering
    \includegraphics[keepaspectratio=true,scale=0.5]{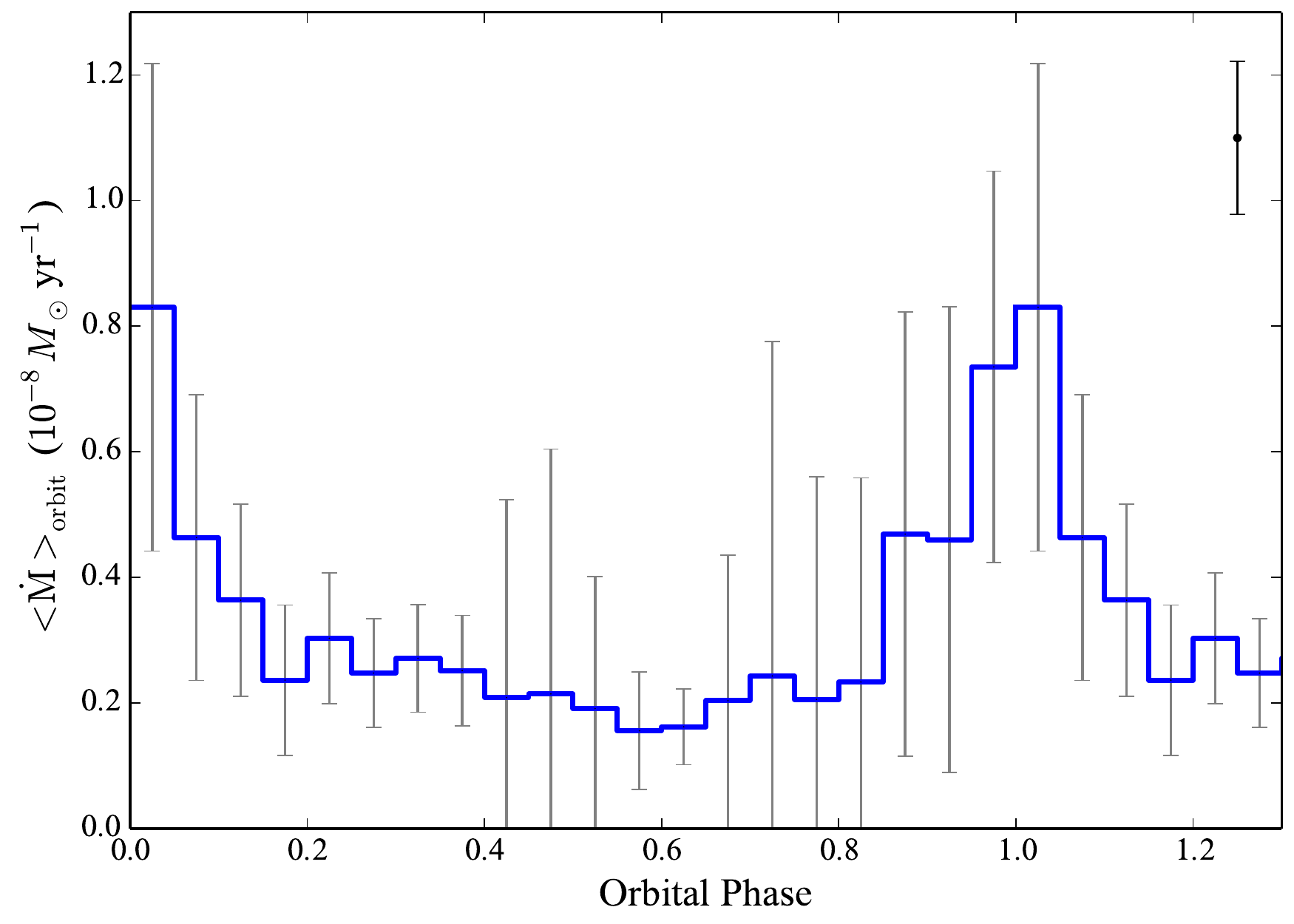}
    \caption{Orbit-averaged (median) mass accretion rate from 10 orbital
      cycles of observation. Error bars are the standard deviation within each
      phase bin. Black error bar in the top right corner denotes the
      propagation of the systematic error of our photometric calibration.}
    \label{fig:lcPAA}
\end{figure}

In addition to the presence of enhanced periastron accretion, the morphology
and timing of the observed accretion events also provide a test of numerical
simulations. Given that large variability exists from orbit to orbit we create
an orbit-averaged accretion rate as a function of orbital phase. First, as to
not over-weight the orbital periods with high-cadence observations, while
still making use of the morphological information they provide, a linear
interpolation of the mass accretion rate is computed and re-sampled at our
average moderate-cadence rate (20 times per orbital period). The median value
from 10 orbital periods is then calculated in phase bins of $\phi=0.05$ (10
measurements per bin) resulting in the orbit average accretion event profile
in Figure \ref{fig:lcPAA}. The error bars at each bin signify the standard
deviation within that bin from orbit-to-orbit. {\it On average}, accretion
rates increase by a factor of $\sim$5 above quiescence at periastron
($\phi=0.95$ to 1.05) with a mostly symmetric rise and decay about periastron.

\begin{figure}[!t]
  \centering
    \includegraphics[keepaspectratio=true,scale=0.5]{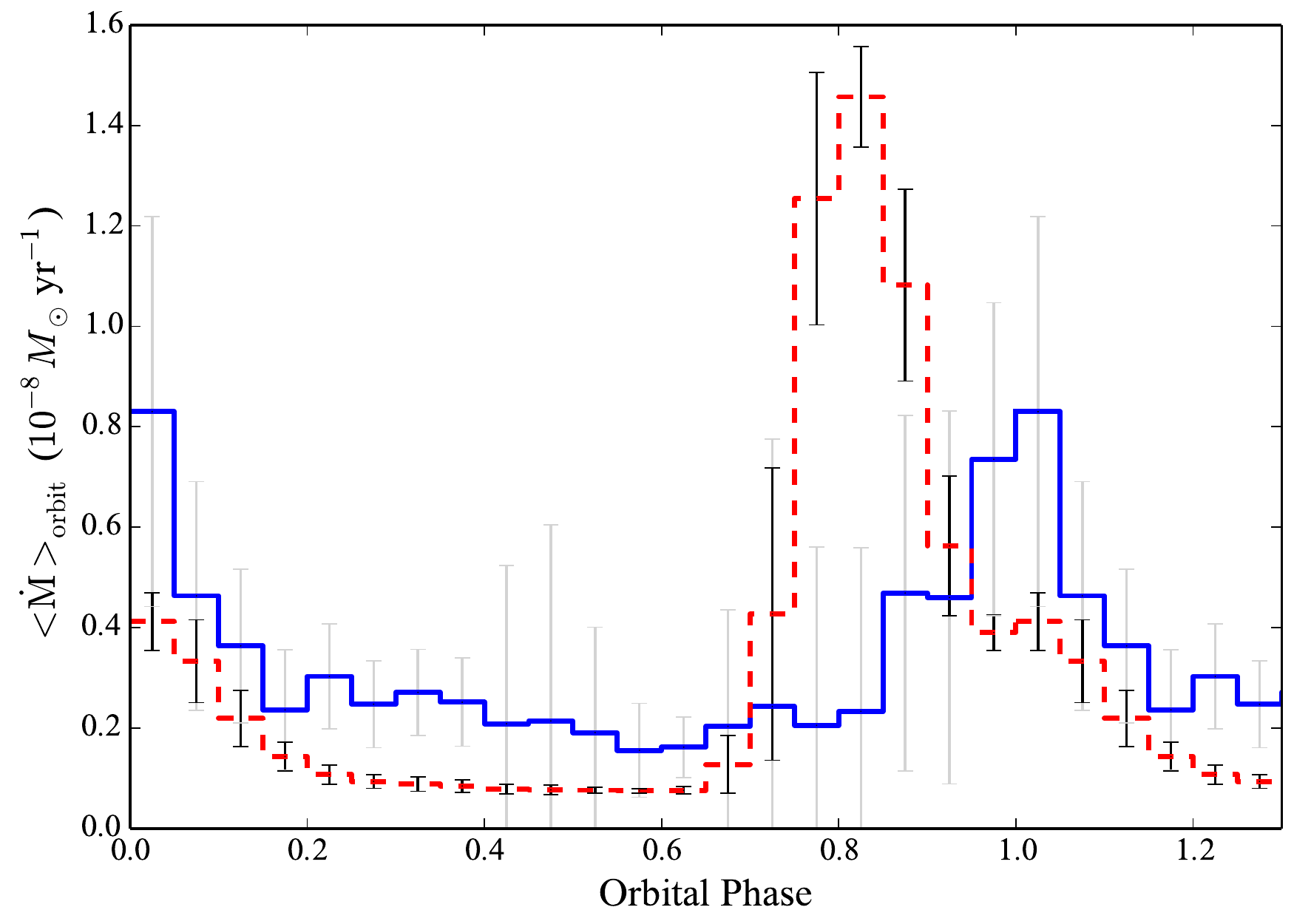}
    \caption{Orbit-averaged (median) mass accretion rate from 10 orbital
      cycles of observation and simulation in blue solid and red dashed lines,
      respectively. Error bars are the standard deviation of the accretion
      rate within each phase bin. Simulation results come from Mu{\~n}oz \&
      Lai (2016) 2D hydrodynamical models. }
    \label{fig:MD}
\end{figure}

To compare our results directly with numerical simulations, we create an orbit
averaged mass accretion rate from the ML2016 2D hydrodynamical models of
binary accretion (D. Mu{\~n}oz, private communication). These models are novel
in that they utilize the adaptive mesh refinement code AREPO
\citep{Springel2010}, extend out to radii of 70$a$, and run for $>$2000
orbital periods reaching full relaxation from the initial conditions out to a
radius of $\sim$5$a$ in the CBD. Using the results from 10 orbital periods of
their scale-free, eccentric ($e=0.5$), equal-mass binary simulation (similar
to DQ Tau; $e=0.568$; $q=0.94$), we perform the same averaging scheme used on
our observations. The simulated accretion rate is normalized by matching the
average accretion rate per orbital period to our observations (Figure
\ref{fig:accretion}). Figure \ref{fig:MD} presents a comparison of the model
and data which, to first order, shows remarkable agreement given the limited
input physics of the model (only gas physics and gravity). Both show
significantly enhanced accretion from $\phi\sim0.8$ to $\sim$1.1.

In detail however, the model and data differ in the specific morphology of the
average accretion event, the orbital phase of peak accretion, and the
consistency of both compared to the observed variability (apparent when
comparing the variability within each phase bin from the model to
observations). Exploring these differences acts to highlight the important
ingredients missing from numerical simulations. In the ML2016 simulations each
star develops a tidally truncated circumstellar disk that extends down to the
stellar radius where mass is deposited. With viscous accretion timescales as
short as 20 orbital periods for disk of this size, circumstellar disks are
replenished each orbital period by a circumbinary accretion stream. This
process acts as an accretion buffer that organizes the incoming material
before it reaches the stars. Bursts of accretion in this case arise not from
accretion stream material impacting the stars themselves but from
companion-induced tidal torques on the circumstellar disks during periastron
approach. These gravitational torques induce non-axisymmetric structures in
the circumstellar disks (spiral arms) that dissipate orbital energy, funneling
material inward.

In the case of DQ Tau however, strong magnetic fields may truncate the inner
edge of the circumstellar disks, potentially to the point that no stable
circumstellar orbits exist. Dynamical outer truncation radii for binary
circumstellar disks are $\sim$0.2$a$ or $\sim$5.6$R_\star$ for DQ Tau's
orbital parameters \citep{Eggleton1983,Miranda&Lai2015}. As discussed above,
the inner magnetospheric truncation radius is likely to vary with the
conditions of incoming material but a typical single-star value is $R_{\rm
  in}\sim5R_\star$, essentially the same as the dynamical truncation. In this
case, the efficency of circumstellar material to buffer accretion streams
would be greatly reduced leaving accretion events more subject to the timing
and extent of material contained within each accretion stream.

This scenario explains the orbit-to-orbit consistency in amplitude and
morphology the ML2016 simulation shows over our observations. The fact that
the simulated accretion rates rise and peak well before ours is likely also
due to the size/existence of circumstellar disks. If the material constituting
DQ Tau's periastron accretion events is provided by the accretion stream of
that orbital period alone, there may be no circumstellar material laying in
wait to be torqued by the companion star, delaying the onset of accretion. In
addition, periastron passages 5 and 7 (Figure \ref{fig:accretion}), for
instance, display discrete accretion events at orbital phases 1.18 and 0.72,
respectively, where companion-induced tidal torques are likely insignificant
given the stellar separation.

It is possible that we have confirmed the observational predictions of
numerical simulations without, necessarily, the same dominant physical
mechanisms at play. Simulations including treatments of magnetism and
radiative transfer may be required for a more in-depth comparison with
short-period systems like DQ Tau. Long-period binaries where the
magnetospheric inner truncation radius is less significant may be well
described by these models.

\begin{figure}[!tb]
  \centering
    \includegraphics[keepaspectratio=true,scale=0.5]{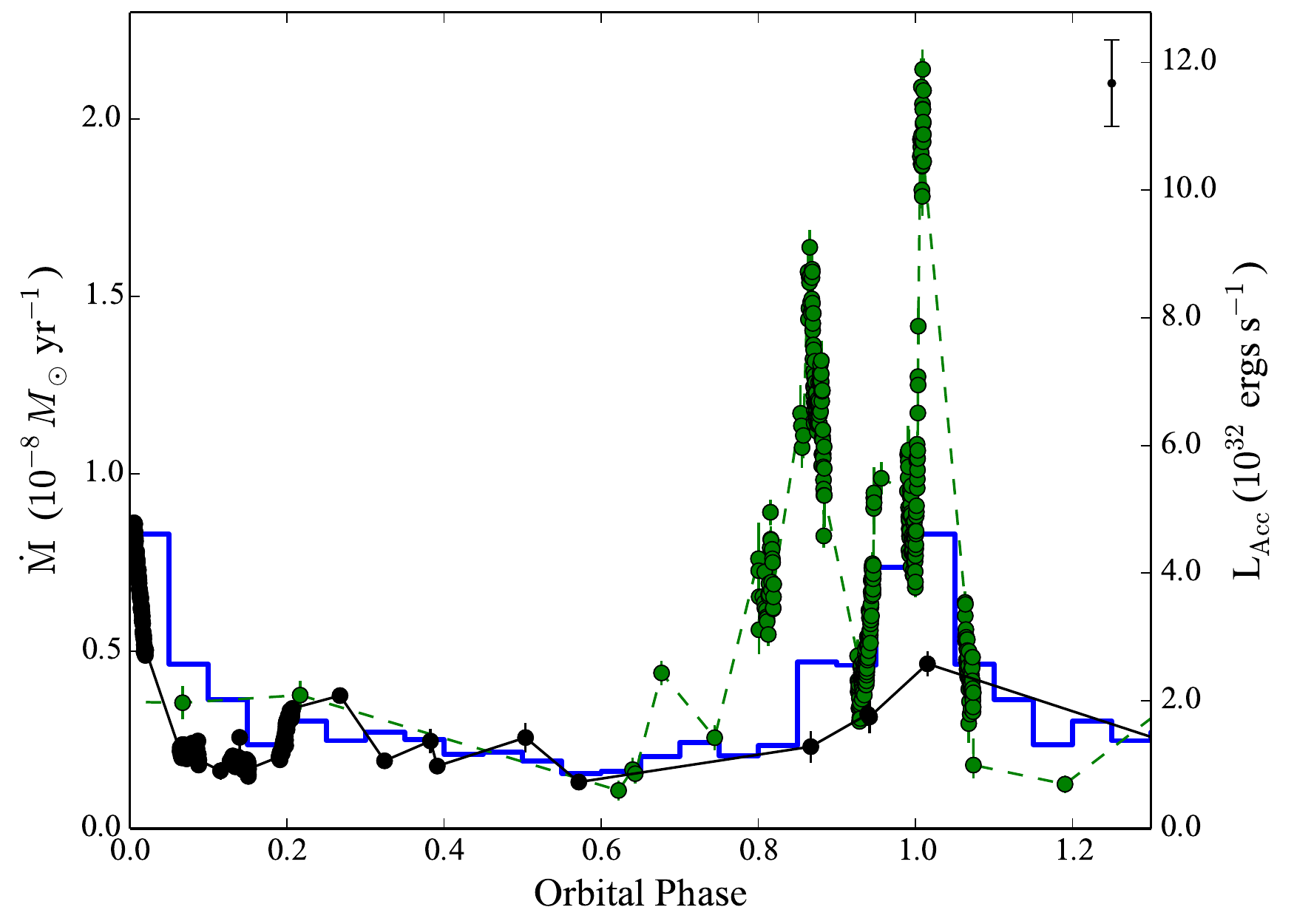}
    \caption{The mass accretion rate (left axis) of the most active and
      passive orbital periods. The green and black data are from orbital
      cycles 2 and 4, respectively. The blue histogram is orbit-averaged
      accretion rate from Figure \ref{fig:lcPAA}. The right axis present the
      corresponding accretion luminosity.}
    \label{fig:M_dot_vary}
\end{figure}

\subsection{Accretion Variability}
\label{variability}

The orbit-averaged accretion rate above provides definitive evidence that
bursts of accretion primarily occur near periastron, consistent with the
predictions of the binary pulsed accretion theory. However, the orbit-averaged
accretion rate provides a very poor description of the behavior in a given
orbit. Figure \ref{fig:M_dot_vary} highlights this variability, presenting one
of the more active and passive orbital periods.

Other than occurring primarily near periastron, accretion events vary in
amplitude, duration, and morphology. Our high-cadence observations reveal
that, rather than a single rise and decay across periastron, accretion occurs
in discrete, short-lived events (Figure \ref{fig:M_dot_vary}). In some sense,
this behavior is not surprising given the large amount of variability seen in
single CTTSs \citep{Rucinskietal2008,Codyetal2014,Staufferetal2014}.  The
\citet{Kulkarni&Romanova2008} 3D MHD simulations of Rayleigh-Taylor unstable
accretion, for instance, provide a good qualitative match to the bursty and
quasi-period nature of accretion on single CTTSs.

Inspection of the bottom panel of Figure \ref{fig:accretion} shows a factor of
$\sim$5 variability (min-to-max) in the mass accreted per orbital period. For
reference, the ML2016 simulation only vary by $\sim$10\% from
orbit-to-orbit. The source of this variability must come from either changes
in the amount of CBD material supplied from one orbit to the next or changes
in the efficiency at which the stars drain their reservoirs of material. If we
assume the amount of material brought in through accretion streams is the same
for every orbit and only the efficiency at which the stars accrete changes, we
would expect orbital periods with low accretion to be followed by those with
high accretion, fueled by ``leftover'' material. Although only 10 orbital
cycles are observed, there does not appear to be any obvious connection
between the mass accreted from one orbital period to the next.

If instead, each star accretes a majority of its bound material within an
orbital period (the case if little/no stable circumstellar material exists),
variability in the mass accreted per orbit would reflect variability in the
mass supplied by the circumbinary streams. The time-variable nature of
gravitational perturbations from an eccentric orbit creates a dynamic and
unstable region near the CBD edge that could supply the inhomogeneities
required to explain our observations. The ML2016 CBDs, for instance, develop
asymmetries that precess around the central gap as well as over-densities that
grow, becoming unstable, and fall inward.

While changes in the stellar accretion efficiency and stream mass are likely
both at play, we find the observed variability is most easily explained by
assuming a significant portion of the circumstellar material is truncated near
the star by magnetic fields, greatly inhibiting the ability to buffer, or
hide, variability in accretion streams. This is supported by the variability
in the accreted mass from orbit-to-orbit as well as the bursty and varied
orbital phases of the near-periastron accretion events. The discrete nature of
the observed accretion events (Figure \ref{fig:M_dot_vary}) may also indicate
an inhomogeneous nature to the material within a given stream that provides a
non-steady flow of material to the stellar surface(s).

Changes in the magnetic field topology almost certainly plays a role in
accretion variability as well. With large-scale magnetic reconnection events
and time-variable ram pressure from accreting material, the state of the
magnetic fields is largely unknown. We find it unlikely that the magnetic
field alone could be responsible for suppressing the accretion rate to the
degree that is observed in some orbital cycles, but it may affect the ability
of the stars to capture stream material, alter the efficiency at which they
drain the reservoir of circumstellar material, and foster the bursty nature of
the observed accretion.

\begin{figure}[!tb]
  \centering
  \includegraphics[keepaspectratio=true,scale=0.5]{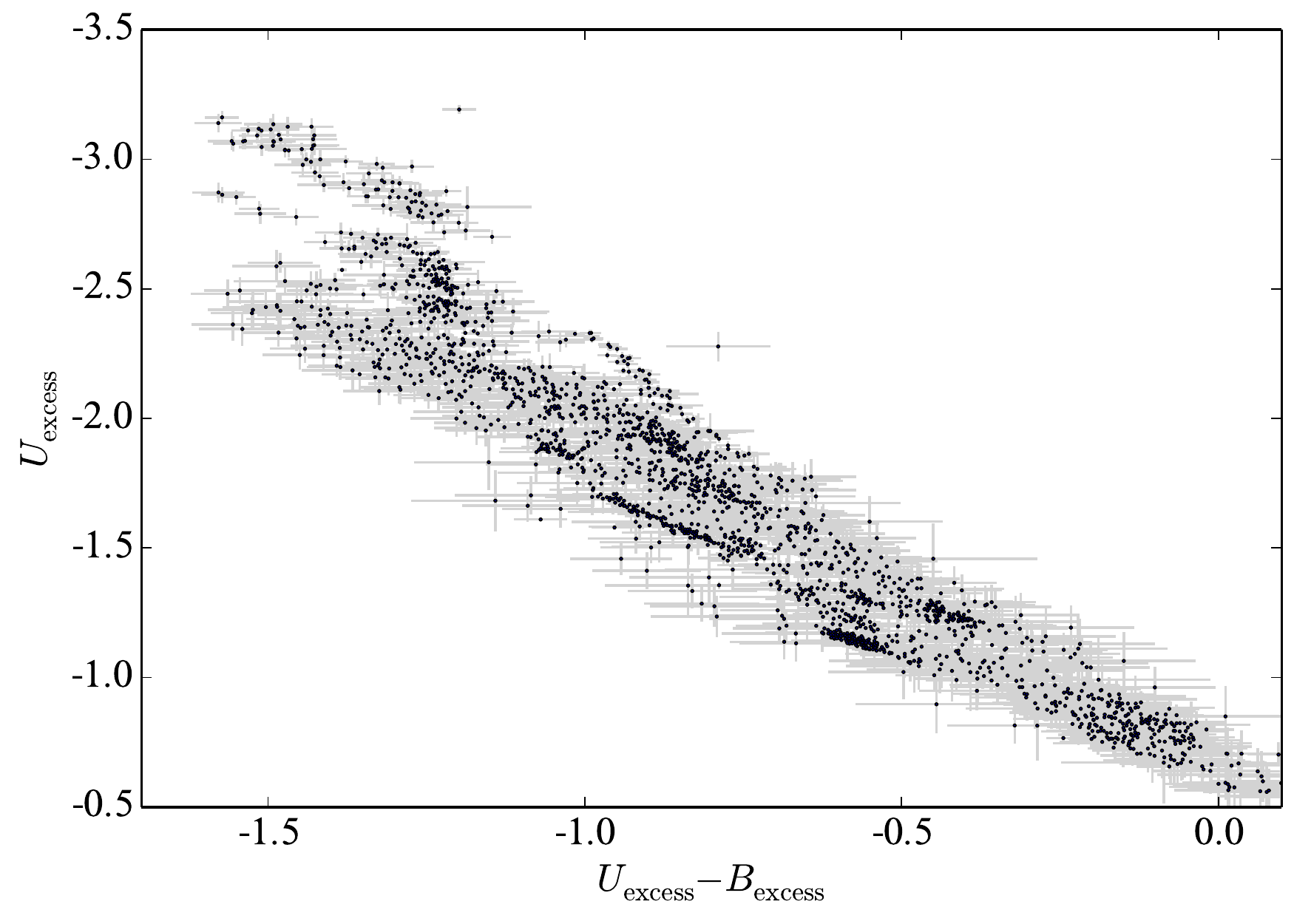}
  \caption{$U$--$B$ color of the excess accretion luminosity versus the
    $U$-band excess above the stellar photosphere (flares removed). The
    $U$-band excess luminosity is a proxy for the mass accretion rate. For
    given $U$-band excess a variety of $U-B$ excess colors are present.}
  \label{fig:UB}
\end{figure}

In addition to variability in the accretion rate itself, the spectral
characteristics of the accretion luminosity is also variable. Figure
\ref{fig:UB} displays the color-magnitude diagram of the $U$-band excess
emission versus the $U$--$B$ excess color. Here we see complex behavior where
the $U$--$B$ excess color is not simply a function of the $U$-band excess, a
proxy for the mass accretion rate. For a given $U$-band excess, a wide range
of $U$--$B$ excess colors exist pointing to different physical conditions of
emitting material for a single $\dot{M}$. Tightly grouped streaks in the
color-magnitude diagram correspond to individual nights of high-cadence
observation where we can see changes in not only the accretion rate, but also
the conditions of accretion.

\citet{Calvet&Gullbring1998} have modeled the emission of accreting CTTSs in
the magnetic paradigm where, for a given mass and stellar radius, the emergent
emission from an accretion column is set by its energy flux, $\rho v^3/2$, and
surface filling factor (also see \citealt{Inglebyetal2013}). Increasing the
energy flux of these models corresponds to an increase in the total emission,
specifically blueward of the Balmer jump which, centered in the $U$-band, is
likely the dominant source of $U-B$ excess color variability. Physically, this
would require either a change in the density of the accreting material, its
velocity, or the size of the accretion site. All three are likely to be
changing in DQ Tau. Inhomogeneities in accretion streams could affect the
density of incoming material while simultaneously compressing the magnetic
field to small $R_{\rm in}$ values which would correspond to small free-fall
velocities (Equation \ref{eqn:MA}). This variable accretion scenario is also
likely to form Rayleigh-Taylor instabilities leading to unstable accretion
flows which can increase the covering fraction of accretion sites
\citep{Kulkarni&Romanova2008}. It is also possible that both stars are
accreting simultaneously under different conditions.

While our four-color photometry does not provide the spectral leverage to
estimate changes in the energy flux or physical size of accretion sites, we
note that when comparing the slope of the $U$-band lightcurve to the excess
$U$--$B$ color, the rise of accretion events are consistently bluer that the
decay.  We interpret the bluer color as a larger emission blueward of the
Balmer jump in the accretion spectrum, corresponding to a higher energy
flux. This behavior suggests that the energy flux is higher at the onset of
accretion events than during their decay.

\subsection{Apastron Accretion Events}
\label{apastacc}

Outside of the predicted periastron accretion events, bursts of accretion also
occur near periastron. This behavior was first observed in DQ Tau by
\citet{Bary&Petersen2014} and is not predicted by any models of eccentric
binary accretion. Apastron events are less visually apparent in the lightcurve
(Figure \ref{fig:accretion}) than periastron events, but are present at a
level capable of producing statistically significant periodicity at twice the
orbital frequency (see Figure \ref{fig:lcPAA} and Section
\ref{PEA}). Prominent examples can be seen at orbital cycles 6.5, 8.5, and 9.5
(Figure \ref{fig:accretion}). While only three strong apastron events are
seen, all three preceed some, but not all, of the periastron passages with
large integrated mass accretion.

We speculate that the source of the apastron events are either ``leftovers''
from inefficient draining during the preceeding orbital cycle, or direct
accretion from CBD. In the ML2016 simulations, each star passes through the
remnants of their companion's unbound accretion stream near apastron that,
without a buffering circumstellar disk, could lead to an accretion
event. Alternatively, asymmetries in the CBD gap may also place material in
the orbital path of the stars leading to direct accretion. If this scenario
were the case, it might explain why subsequent periastron accretion events are
large. A favorable alignment of the orientation of a CBD asymmetry at apastron
passage might produce an apastron accretion event while placing more material
than average under the gravitational influence of the star resulting in a
larger accretion stream for the ensuing periastron.

\section{SUMMARY \& CONCLUSIONS}
\label{conc}

With moderate-cadence photometry from LCOGT, supplemented with high-cadence
photometry from the WIYN 0.9m and ARCSAT telescopes, we have obtained a
comprehensive data set capable of characterizing variability and its physical
mechanism in the T Tauri binary DQ Tau. Critically, our observations combine
multi-orbit coverage, the time-resolution necessary to distinguish stellar
flares from accretion variability, and $U$-band photometry capable of
determining accretion rates.

Analysis of the lightcurve morphology reveals few events that resemble the
characteristic shape of stellar flares. We develop a flare finding scheme
aimed at detecting impulsive brightening events based on the characteristics
of M dwarf flares that are then visually inspected. Two flares are identified,
one classical and one gradual/slow above an average detection threshold of
$\Delta U=0.32$ mag. Modeling the classical flare with the
\citet{Davenportetal2014} template places its integrated energy in good
agreement with flares observed on other pre-MS stars. We find that optical
flares are responsible for a very small portion of the optical variability,
occurring in $\sim3$\% of our high-cadence coverage.

Under the assumption that the optical counterpart to the large mm-wave flares
observed by \citet{Salteretal2010} resemble those of active M dwarfs, we
further conclude that magnetic reconnection events from colliding
magnetospheres do not have a significant effect on the optical
lightcurve. With the site of energy generation in these events occurring far
from the stellar surfaces ($\sim6R_\star$), the transport of energy to the
photosphere to create an optical counterpart (the classical solar/stellar
flare scenario) is complex and may suffer from confinement and energy
losses. Even if that energy were deposited efficiently in the the stellar
surface, the predicted energy budget from colliding magnetospheres is a factor
of 10$^3$ less than the observed optical output near periastron.  The two
flares events that are found are in all likelihood magnetic reconnection
events in a single magnetosphere near the stellar surface.

Removing the contribution from flares, we characterize the accretion
variability in DQ Tau by converting the $U$-band excess luminosity into an
accretion rate. Statistically significant periodicity in the mass accretion
rate is present at the orbital period, powered by consistent periastron
accretion events, that confirms the theoretical prediction of accretion in
eccentric binaries. During some orbits, 90\% of the mass accreted in that
orbital period occurs near periastron ($\phi$=0.7-1.3). We determine the
median accretion rate as a function of orbital phase to characterize the
average morphology and amplitude of accretion events. On average, accretion
rates increase by a factor of 5 near periastron. This result is in good
agreement with the \citet{Munoz&Lai2016} hydrodynamical models.

Moving beyond the orbit-averaged accretion rate, we find complex variability
from one orbital passage to the next. Broadly speaking, the results of
hydrodynamic simulations match our observations, supporting the picture that
streams of circumbinary disk (CBD) material are periodically brought into the
central gap that feed accretion events near periastron. In detail however, the
way in which these flows interact with the stars is more complex than the
models depict. The scale of DQ Tau's orbit results in a close match between
inner and outer truncation radii of a circumstellar disk; the inner set by the
stellar magnetosphere and the outer set by orbital resonances. The lack of
extensive, stable circumstellar disks around the DQ Tau primary and secondary
leaves accretion responsive to variability in the streams themselves and
therefore the CBD. A picture emerges of inhomogeneity at the inner edge of the
CBD providing streams to the central binary that are variable in mass from one
orbit to the next, and streams that are non-steady or discrete in
nature. These inhomogeneities translate into variations in the amount and
timing of material accreted per orbital period and the discrete, bursty nature
of the observed accretion events. Variability in the spectral characteristics
of the accretion events reveal changes in the combined density and velocity
(energy flux) of accretion flows as well as physical size of the accretion
column. We attribute this behavior as changes in the characteristics of the
accretion streams and their impact on the topology of the stellar magnetic
fields.

Quasi-periodic accretion events near apastron are also observed. Elevated
apastron accretion has been detected in DQ Tau previously
\citep{Bary&Petersen2014}, but this is the first time in which these events
are seen to be (quasi-)periodic in nature. In general, they occur less
frequently and at smaller amplitudes when compared to periastron
accretion. Although apastron accretion events are not predicted by the binary
accretion theory, they may be a unique feature of very-short-period, eccentric
binaries where the absence of stable circumstellar material leads to direct
accretion of unbound material within the CBD gap or from CBD material itself
in the orbital path.

While confronting the complex nature of binary accretion is daunting from both
an observational and theoretical perspective, efforts to characterize these
types of systems have far-reaching implications for accretion, disk physics,
binary stellar evolution, and planet formation in the binary environment.

\acknowledgments The authors would like to thank Diego Mu{\~n}oz and Dong Lai
for providing the results of their simulations and for many useful
discussions. We thank Suzanne Hawley and the Astrophysical Research Consortium
(ARC) for their generous allocation of ARCSAT commissioning time as well as
Flynn Hasse and the WIYN 0.9m synoptic observers, Stephen Gilliam,
Fran\c{c}ois Dufour, and William Romanishin. This work makes use of
observations from the LCOGT network and observations obtained with Apache
Point Observatory's 0.5-m Astrophysical Research Consortium Small Aperture
Telescope. B.M.T acknowledges support from a Sigma Xi Grant-in-Aid of Research
and the University of Wisconsin--Madison Graduate School.


\begin{thebibliography}{86}
\bibitem[{{Adams} {et~al.}(2011){Adams}, {Cai}, {Galli}, {Lizano}, \&
  {Shu}}]{Adamsetal2011}
{Adams}, F.~C., {Cai}, M.~J., {Galli}, D., {Lizano}, S., \& {Shu}, F.~H. 2011,
  \apj, 743, 175

\bibitem[{{Alcal{\'a}} {et~al.}(2014){Alcal{\'a}}, {Natta}, {Manara}, {Spezzi},
  {Stelzer}, {Frasca}, {Biazzo}, {Covino}, {Randich}, {Rigliaco}, {Testi},
  {Comer{\'o}n}, {Cupani}, \& {D'Elia}}]{Alcalaetal2014}
{Alcal{\'a}}, J.~M., {Natta}, A., {Manara}, C.~F., {et~al.} 2014, \aap, 561, A2

\bibitem[{{Alencar} {et~al.}(2010){Alencar}, {Teixeira}, {Guimar{\~a}es},
  {McGinnis}, {Gameiro}, {Bouvier}, {Aigrain}, {Flaccomio}, \&
  {Favata}}]{Alencaretal2010}
{Alencar}, S.~H.~P., {Teixeira}, P.~S., {Guimar{\~a}es}, M.~M., {et~al.} 2010,
  \aap, 519, A88

\bibitem[{{Alexander} {et~al.}(2014){Alexander}, {Pascucci}, {Andrews},
  {Armitage}, \& {Cieza}}]{Alexanderetal2014}
{Alexander}, R., {Pascucci}, I., {Andrews}, S., {Armitage}, P., \& {Cieza}, L.
  2014, Protostars and Planets VI, 475

\bibitem[{{Allred} {et~al.}(2006){Allred}, {Hawley}, {Abbett}, \&
  {Carlsson}}]{Allredetal2006}
{Allred}, J.~C., {Hawley}, S.~L., {Abbett}, W.~P., \& {Carlsson}, M. 2006,
  \apj, 644, 484

\bibitem[{{Andrews} {et~al.}(2011){Andrews}, {Wilner}, {Espaillat}, {Hughes},
  {Dullemond}, {McClure}, {Qi}, \& {Brown}}]{Andrewsetal2011}
{Andrews}, S.~M., {Wilner}, D.~J., {Espaillat}, C., {et~al.} 2011, \apj, 732,
  42

\bibitem[{{Artymowicz} \& {Lubow}(1994)}]{Artymowicz&Lubow1994}
{Artymowicz}, P., \& {Lubow}, S.~H. 1994, \apj, 421, 651

\bibitem[{{Artymowicz} \& {Lubow}(1996)}]{Artymowicz&Lubow1996}
---. 1996, \apjl, 467, L77

\bibitem[{{Aschwanden} {et~al.}(1998){Aschwanden}, {Schwartz}, \&
  {Dennis}}]{Aschwandenetal1998}
{Aschwanden}, M.~J., {Schwartz}, R.~A., \& {Dennis}, B.~R. 1998, \apj, 502, 468

\bibitem[{{Baraffe} {et~al.}(2015){Baraffe}, {Homeier}, {Allard}, \&
  {Chabrier}}]{Baraffeetal2015}
{Baraffe}, I., {Homeier}, D., {Allard}, F., \& {Chabrier}, G. 2015, \aap, 577,
  A42

\bibitem[{{Bary} \& {Petersen}(2014)}]{Bary&Petersen2014}
{Bary}, J.~S., \& {Petersen}, M.~S. 2014, \apj, 792, 64

\bibitem[{{Basri} {et~al.}(1997){Basri}, {Johns-Krull}, \&
  {Mathieu}}]{Basrietal1997}
{Basri}, G., {Johns-Krull}, C.~M., \& {Mathieu}, R.~D. 1997, \aj, 114, 781

\bibitem[{{Beck} {et~al.}(2012){Beck}, {Bary}, {Dutrey}, {Pi{\'e}tu},
  {Guilloteau}, {Lubow}, \& {Simon}}]{Becketal2012}
{Beck}, T.~L., {Bary}, J.~S., {Dutrey}, A., {et~al.} 2012, \apj, 754, 72

\bibitem[{Bertin \& Arnouts(1996)}]{Bertin&Arnouts1996}
Bertin, E., \& Arnouts, S. 1996, Astron. Astrophys. Suppl. Ser., 117, 393

\bibitem[{{Boden} {et~al.}(2009){Boden}, {Akeson}, {Sargent}, {Carpenter},
  {Ciardi}, {Bary}, \& {Skrutskie}}]{Bodenetal2009}
{Boden}, A.~F., {Akeson}, R.~L., {Sargent}, A.~I., {et~al.} 2009, \apjl, 696,
  L111

\bibitem[{{Bouvier} {et~al.}(1995){Bouvier}, {Covino}, {Kovo}, {Martin},
  {Matthews}, {Terranegra}, \& {Beck}}]{Bouvieretal1995}
{Bouvier}, J., {Covino}, E., {Kovo}, O., {et~al.} 1995, \aap, 299, 89

\bibitem[{{Brown}(1971)}]{Brown1971}
{Brown}, J.~C. 1971, \solphys, 18, 489

\bibitem[{{Brown} {et~al.}(2013){Brown}, {Baliber}, {Bianco}, {Bowman},
  {Burleson}, {Conway}, {Crellin}, {Depagne}, {De Vera}, {Dilday}, {Dragomir},
  {Dubberley}, {Eastman}, {Elphick}, {Falarski}, {Foale}, {Ford}, {Fulton},
  {Garza}, {Gomez}, {Graham}, {Greene}, {Haldeman}, {Hawkins}, {Haworth},
  {Haynes}, {Hidas}, {Hjelstrom}, {Howell}, {Hygelund}, {Lister}, {Lobdill},
  {Martinez}, {Mullins}, {Norbury}, {Parrent}, {Paulson}, {Petry}, {Pickles},
  {Posner}, {Rosing}, {Ross}, {Sand}, {Saunders}, {Shobbrook}, {Shporer},
  {Street}, {Thomas}, {Tsapras}, {Tufts}, {Valenti}, {Vander Horst}, {Walker},
  {White}, \& {Willis}}]{Brownetal2013}
{Brown}, T.~M., {Baliber}, N., {Bianco}, F.~B., {et~al.} 2013, \pasp, 125, 1031

\bibitem[{{Butler} {et~al.}(2015){Butler}, {Erkan}, {Budding}, {Doyle},
  {Foing}, {Bromage}, {Kellett}, {Frueh}, {Huovelin}, {Brown}, \&
  {Neff}}]{Butleretal2015}
{Butler}, C.~J., {Erkan}, N., {Budding}, E., {et~al.} 2015, \mnras, 446, 4205

\bibitem[{{Calvet} \& {Gullbring}(1998)}]{Calvet&Gullbring1998}
{Calvet}, N., \& {Gullbring}, E. 1998, \apj, 509, 802

\bibitem[{{Cardelli} {et~al.}(1989){Cardelli}, {Clayton}, \&
  {Mathis}}]{Cardellietal1989}
{Cardelli}, J.~A., {Clayton}, G.~C., \& {Mathis}, J.~S. 1989, \apj, 345, 245

\bibitem[{{Carr} {et~al.}(2001){Carr}, {Mathieu}, \& {Najita}}]{Carretal2001}
{Carr}, J.~S., {Mathieu}, R.~D., \& {Najita}, J.~R. 2001, \apj, 551, 454

\bibitem[{{Cody} {et~al.}(2014){Cody}, {Stauffer}, {Baglin}, {Micela},
  {Rebull}, {Flaccomio}, {Morales-Calder{\'o}n}, {Aigrain}, {Bouvier},
  {Hillenbrand}, {Gutermuth}, {Song}, {Turner}, {Alencar}, {Zwintz},
  {Plavchan}, {Carpenter}, {Findeisen}, {Carey}, {Terebey}, {Hartmann},
  {Calvet}, {Teixeira}, {Vrba}, {Wolk}, {Covey}, {Poppenhaeger}, {G{\"u}nther},
  {Forbrich}, {Whitney}, {Affer}, {Herbst}, {Hora}, {Barrado}, {Holtzman},
  {Marchis}, {Wood}, {Medeiros Guimar{\~a}es}, {Lillo Box}, {Gillen},
  {McQuillan}, {Espaillat}, {Allen}, {D'Alessio}, \& {Favata}}]{Codyetal2014}
{Cody}, A.~M., {Stauffer}, J., {Baglin}, A., {et~al.} 2014, \aj, 147, 82

\bibitem[{Czekala {et~al.}(2016)Czekala, Andrews, Torres, Jensen, Stassun,
  Wilner, \& Latham}]{Czekalaetal2016}
Czekala, I., Andrews, S.~M., Torres, G., {et~al.} 2016, Astrophys. J., 818, 156

\bibitem[{{Dal} \& {Evren}(2010)}]{Dal&Evren2010}
{Dal}, H.~A., \& {Evren}, S. 2010, \aj, 140, 483

\bibitem[{Davenport {et~al.}(2014)Davenport, Hawley, Hebb, Wisniewski,
  Kowalski, Johnson, Malatesta, Peraza, Keil, Silverberg, Jansen, Scheffler,
  Berdis, Larsen, \& Hilton}]{Davenportetal2014}
Davenport, J. R.~A., Hawley, S.~L., Hebb, L., {et~al.} 2014, Astrophys. J.,
  797, 122

\bibitem[{de~Val-Borro {et~al.}(2011)de~Val-Borro, Gahm, Stempels, \&
  Pepliński}]{DeVal-Borroetal2011}
de~Val-Borro, M., Gahm, G.~F., Stempels, H.~C., \& Pepliński, A. 2011, Mon.
  Not. R. Astron. Soc., 413, 2679

\bibitem[{{Eggleton}(1983)}]{Eggleton1983}
{Eggleton}, P.~P. 1983, \apj, 268, 368

\bibitem[{{Fern{\'a}ndez} {et~al.}(2004){Fern{\'a}ndez}, {Stelzer}, {Henden},
  {Grankin}, {Gameiro}, {Costa}, {Guenther}, {Amado}, \&
  {Rodriguez}}]{Fernandezetal2004}
{Fern{\'a}ndez}, M., {Stelzer}, B., {Henden}, A., {et~al.} 2004, \aap, 427, 263

\bibitem[{{Fletcher} {et~al.}(2011){Fletcher}, {Dennis}, {Hudson}, {Krucker},
  {Phillips}, {Veronig}, {Battaglia}, {Bone}, {Caspi}, {Chen}, {Gallagher},
  {Grigis}, {Ji}, {Liu}, {Milligan}, \& {Temmer}}]{Fletcheretal2011}
{Fletcher}, L., {Dennis}, B.~R., {Hudson}, H.~S., {et~al.} 2011, \ssr, 159, 19

\bibitem[{{Frescura} {et~al.}(2008){Frescura}, {Engelbrecht}, \&
  {Frank}}]{Frescuraetal2008}
{Frescura}, F.~A.~M., {Engelbrecht}, C.~A., \& {Frank}, B.~S. 2008, \mnras,
  388, 1693

\bibitem[{{Gahm}(1990)}]{Gahm1990}
{Gahm}, G.~F. 1990, in IAU Symposium, Vol. 137, Flare Stars in Star Clusters,
  Associations and the Solar Vicinity, ed. L.~V. {Mirzoian}, B.~R. {Pettersen},
  \& M.~K. {Tsvetkov}, 193--206

\bibitem[{{G{\'o}mez de Castro} {et~al.}(2013){G{\'o}mez de Castro},
  {L{\'o}pez-Santiago}, {Talavera}, {Sytov}, \&
  {Bisikalo}}]{GomezdeCastroetal2013}
{G{\'o}mez de Castro}, A.~I., {L{\'o}pez-Santiago}, J., {Talavera}, A.,
  {Sytov}, A.~Y., \& {Bisikalo}, D. 2013, \apj, 766, 62

\bibitem[{{Grankin} {et~al.}(2008){Grankin}, {Bouvier}, {Herbst}, \&
  {Melnikov}}]{Grankinetal2008}
{Grankin}, K.~N., {Bouvier}, J., {Herbst}, W., \& {Melnikov}, S.~Y. 2008, \aap,
  479, 827

\bibitem[{{Gullbring} {et~al.}(1998){Gullbring}, {Hartmann}, {Briceno}, \&
  {Calvet}}]{Gullbringetal1998}
{Gullbring}, E., {Hartmann}, L., {Briceno}, C., \& {Calvet}, N. 1998, \apj,
  492, 323

\bibitem[{{G{\"u}nther} \& {Kley}(2002)}]{Gunther&Kley2002}
{G{\"u}nther}, R., \& {Kley}, W. 2002, \aap, 387, 550

\bibitem[{{Harris} {et~al.}(2012){Harris}, {Andrews}, {Wilner}, \&
  {Kraus}}]{Harrisetal2012}
{Harris}, R.~J., {Andrews}, S.~M., {Wilner}, D.~J., \& {Kraus}, A.~L. 2012,
  \apj, 751, 115

\bibitem[{{Hartmann} {et~al.}(1994){Hartmann}, {Hewett}, \&
  {Calvet}}]{Hartmannetal1994}
{Hartmann}, L., {Hewett}, R., \& {Calvet}, N. 1994, \apj, 426, 669

\bibitem[{Hawley {et~al.}(2014)Hawley, Davenport, Kowalski, Wisniewski, Hebb,
  Deitrick, \& Hilton}]{Hawleyetal2014}
Hawley, S.~L., Davenport, J. R.~A., Kowalski, A.~F., {et~al.} 2014, Astrophys.
  J., 797, 121

\bibitem[{{Hawley} \& {Pettersen}(1991)}]{Hawley&Pettersen1991}
{Hawley}, S.~L., \& {Pettersen}, B.~R. 1991, \apj, 378, 725

\bibitem[{{Herczeg} \& {Hillenbrand}(2008)}]{Herczeg&Hillenbrand2008}
{Herczeg}, G.~J., \& {Hillenbrand}, L.~A. 2008, \apj, 681, 594

\bibitem[{{Herczeg} \& {Hillenbrand}(2014)}]{Herczeg&Hillenbrand2014}
---. 2014, \apj, 786, 97

\bibitem[{{Honeycutt}(1992)}]{Honeycutt1992}
{Honeycutt}, R.~K. 1992, \pasp, 104, 435

\bibitem[{{Ingleby} {et~al.}(2015){Ingleby}, {Espaillat}, {Calvet}, {Sitko},
  {Russell}, \& {Champney}}]{Inglebyetal2015}
{Ingleby}, L., {Espaillat}, C., {Calvet}, N., {et~al.} 2015, \apj, 805, 149

\bibitem[{{Ingleby} {et~al.}(2011){Ingleby}, {Calvet}, {Bergin}, {Herczeg},
  {Brown}, {Alexander}, {Edwards}, {Espaillat}, {France}, {Gregory},
  {Hillenbrand}, {Roueff}, {Valenti}, {Walter}, {Johns-Krull}, {Brown},
  {Linsky}, {McClure}, {Ardila}, {Abgrall}, {Bethell}, {Hussain}, \&
  {Yang}}]{Inglebyetal2011}
{Ingleby}, L., {Calvet}, N., {Bergin}, E., {et~al.} 2011, \apj, 743, 105

\bibitem[{{Ingleby} {et~al.}(2013){Ingleby}, {Calvet}, {Herczeg}, {Blaty},
  {Walter}, {Ardila}, {Alexander}, {Edwards}, {Espaillat}, {Gregory},
  {Hillenbrand}, \& {Brown}}]{Inglebyetal2013}
{Ingleby}, L., {Calvet}, N., {Herczeg}, G., {et~al.} 2013, \apj, 767, 112

\bibitem[{{Jensen} \& {Mathieu}(1997)}]{Jensen&Mathieu1997}
{Jensen}, E.~L.~N., \& {Mathieu}, R.~D. 1997, \aj, 114, 301

\bibitem[{{Jensen} {et~al.}(1996){Jensen}, {Mathieu}, \&
  {Fuller}}]{Jensenetal1996}
{Jensen}, E.~L.~N., {Mathieu}, R.~D., \& {Fuller}, G.~A. 1996, \apj, 458, 312

\bibitem[{{Jester} {et~al.}(2005){Jester}, {Schneider}, {Richards}, {Green},
  {Schmidt}, {Hall}, {Strauss}, {Vanden Berk}, {Stoughton}, {Gunn},
  {Brinkmann}, {Kent}, {Smith}, {Tucker}, \& {Yanny}}]{Jesteretal2005}
{Jester}, S., {Schneider}, D.~P., {Richards}, G.~T., {et~al.} 2005, \aj, 130,
  873

\bibitem[{{Johns-Krull}(2007)}]{Johns-Krull2007}
{Johns-Krull}, C.~M. 2007, \apj, 664, 975

\bibitem[{{Johnstone} {et~al.}(2014){Johnstone}, {Jardine}, {Gregory},
  {Donati}, \& {Hussain}}]{Johnstoneetal2014}
{Johnstone}, C.~P., {Jardine}, M., {Gregory}, S.~G., {Donati}, J.-F., \&
  {Hussain}, G. 2014, \mnras, 437, 3202

\bibitem[{{Kenyon} {et~al.}(1994){Kenyon}, {Dobrzycka}, \&
  {Hartmann}}]{Kenyonetal1994}
{Kenyon}, S.~J., {Dobrzycka}, D., \& {Hartmann}, L. 1994, \aj, 108, 1872

\bibitem[{{Kenyon} \& {Hartmann}(1995)}]{Kenyon&Hartmann1995}
{Kenyon}, S.~J., \& {Hartmann}, L. 1995, \apjs, 101, 117

\bibitem[{{Koen}(2015)}]{Koen2015}
{Koen}, C. 2015, \mnras, 449, 1704

\bibitem[{{Kowalski} {et~al.}(2015){Kowalski}, {Hawley}, {Carlsson}, {Allred},
  {Uitenbroek}, {Osten}, \& {Holman}}]{Kowalskietal2015}
{Kowalski}, A.~F., {Hawley}, S.~L., {Carlsson}, M., {et~al.} 2015, \solphys,
  290, 3487

\bibitem[{Kowalski {et~al.}(2010)Kowalski, Hawley, Holtzman, Wisniewski, \&
  Hilton}]{Kowalskietal2010}
Kowalski, A.~F., Hawley, S.~L., Holtzman, J.~A., Wisniewski, J.~P., \& Hilton,
  E.~J. 2010, Astrophys. J., 714, L98

\bibitem[{{Kowalski} {et~al.}(2013){Kowalski}, {Hawley}, {Wisniewski}, {Osten},
  {Hilton}, {Holtzman}, {Schmidt}, \& {Davenport}}]{Kowalskietal2013}
{Kowalski}, A.~F., {Hawley}, S.~L., {Wisniewski}, J.~P., {et~al.} 2013, \apjs,
  207, 15

\bibitem[{{Kraus} {et~al.}(2011){Kraus}, {Ireland}, {Martinache}, \&
  {Hillenbrand}}]{Krausetal2011}
{Kraus}, A.~L., {Ireland}, M.~J., {Martinache}, F., \& {Hillenbrand}, L.~A.
  2011, \apj, 731, 8

\bibitem[{Kulkarni \& Romanova(2008)}]{Kulkarni&Romanova2008}
Kulkarni, A.~K., \& Romanova, M.~M. 2008, Mon. Not. R. Astron. Soc., 386, 673

\bibitem[{{Lacy} {et~al.}(1976){Lacy}, {Moffett}, \& {Evans}}]{Lacyetal1976}
{Lacy}, C.~H., {Moffett}, T.~J., \& {Evans}, D.~S. 1976, \apjs, 30, 85

\bibitem[{{Ma{\'{\i}}z Apell{\'a}niz}(2006)}]{MaizApellaniz2006}
{Ma{\'{\i}}z Apell{\'a}niz}, J. 2006, \aj, 131, 1184

\bibitem[{{Massi} {et~al.}(2006){Massi}, {Forbrich}, {Menten},
  {Torricelli-Ciamponi}, {Neidh{\"o}fer}, {Leurini}, \&
  {Bertoldi}}]{Massietal2006}
{Massi}, M., {Forbrich}, J., {Menten}, K.~M., {et~al.} 2006, \aap, 453, 959

\bibitem[{{Massi} {et~al.}(2002){Massi}, {Menten}, \&
  {Neidh{\"o}fer}}]{Massietal2002}
{Massi}, M., {Menten}, K., \& {Neidh{\"o}fer}, J. 2002, \aap, 382, 152

\bibitem[{{Mathieu} {et~al.}(1997){Mathieu}, {Stassun}, {Basri}, {Jensen},
  {Johns-Krull}, {Valenti}, \& {Hartmann}}]{Mathieuetal1997}
{Mathieu}, R.~D., {Stassun}, K., {Basri}, G., {et~al.} 1997, \aj, 113, 1841

\bibitem[{{Miranda} \& {Lai}(2015)}]{Miranda&Lai2015}
{Miranda}, R., \& {Lai}, D. 2015, \mnras, 452, 2396

\bibitem[{{Mu{\~n}oz} \& {Lai}(2016)}]{Munoz&Lai2016}
{Mu{\~n}oz}, D.~J., \& {Lai}, D. 2016, \apj, 827, 43

\bibitem[{{Oke} \& {Gunn}(1982)}]{Oke&Gunn1982}
{Oke}, J.~B., \& {Gunn}, J.~E. 1982, \pasp, 94, 586

\bibitem[{{Orlando} {et~al.}(2013){Orlando}, {Bonito}, {Argiroffi}, {Reale},
  {Peres}, {Miceli}, {Matsakos}, {Stehl{\'e}}, {Ibgui}, {de Sa}, {Chi{\`e}ze},
  \& {Lanz}}]{Orlandoetal2013}
{Orlando}, S., {Bonito}, R., {Argiroffi}, C., {et~al.} 2013, \aap, 559, A127

\bibitem[{{Osten} {et~al.}(2005){Osten}, {Hawley}, {Allred}, {Johns-Krull}, \&
  {Roark}}]{Ostenetal2005}
{Osten}, R.~A., {Hawley}, S.~L., {Allred}, J.~C., {Johns-Krull}, C.~M., \&
  {Roark}, C. 2005, \apj, 621, 398

\bibitem[{{Panagi} \& {Andrews}(1995)}]{Panagi&Andrews1995}
{Panagi}, P.~M., \& {Andrews}, A.~D. 1995, \mnras, 277, 423

\bibitem[{{Pearce} \& {Harrison}(1990)}]{Pearce&Harrison1990}
{Pearce}, G., \& {Harrison}, R.~A. 1990, \aap, 228, 513

\bibitem[{{Pickles} \& {Depagne}(2010)}]{Pickles&Depagne2010}
{Pickles}, A., \& {Depagne}, {\'E}. 2010, \pasp, 122, 1437

\bibitem[{{Press} {et~al.}(1992){Press}, {Teukolsky}, {Vetterling}, \&
  {Flannery}}]{Pressetal1992}
{Press}, W.~H., {Teukolsky}, S.~A., {Vetterling}, W.~T., \& {Flannery}, B.~P.
  1992, {Numerical recipes in FORTRAN. The art of scientific computing}

\bibitem[{{Raghavan} {et~al.}(2010){Raghavan}, {McAlister}, {Henry}, {Latham},
  {Marcy}, {Mason}, {Gies}, {White}, \& {ten Brummelaar}}]{Raghavanetal2010}
{Raghavan}, D., {McAlister}, H.~A., {Henry}, T.~J., {et~al.} 2010, \apjs, 190,
  1

\bibitem[{{Rucinski} {et~al.}(2008){Rucinski}, {Matthews}, {Kuschnig},
  {Pojma{\'n}ski}, {Rowe}, {Guenther}, {Moffat}, {Sasselov}, {Walker}, \&
  {Weiss}}]{Rucinskietal2008}
{Rucinski}, S.~M., {Matthews}, J.~M., {Kuschnig}, R., {et~al.} 2008, \mnras,
  391, 1913

\bibitem[{{Salter} {et~al.}(2008){Salter}, {Hogerheijde}, \&
  {Blake}}]{Salteretal2008}
{Salter}, D.~M., {Hogerheijde}, M.~R., \& {Blake}, G.~A. 2008, \aap, 492, L21

\bibitem[{{Salter} {et~al.}(2010){Salter}, {K{\'o}sp{\'a}l}, {Getman},
  {Hogerheijde}, {van Kempen}, {Carpenter}, {Blake}, \&
  {Wilner}}]{Salteretal2010}
{Salter}, D.~M., {K{\'o}sp{\'a}l}, {\'A}., {Getman}, K.~V., {et~al.} 2010,
  \aap, 521, A32

\bibitem[{{Scargle}(1982)}]{Scargle1982}
{Scargle}, J.~D. 1982, \apj, 263, 835

\bibitem[{{Shu} {et~al.}(1994){Shu}, {Najita}, {Ostriker}, {Wilkin}, {Ruden},
  \& {Lizano}}]{Shuetal1994}
{Shu}, F., {Najita}, J., {Ostriker}, E., {et~al.} 1994, \apj, 429, 781

\bibitem[{{Springel}(2010)}]{Springel2010}
{Springel}, V. 2010, \mnras, 401, 791

\bibitem[{{Stauffer} {et~al.}(2014){Stauffer}, {Cody}, {Baglin}, {Alencar},
  {Rebull}, {Hillenbrand}, {Venuti}, {Turner}, {Carpenter}, {Plavchan},
  {Findeisen}, {Carey}, {Terebey}, {Morales-Calder{\'o}n}, {Bouvier}, {Micela},
  {Flaccomio}, {Song}, {Gutermuth}, {Hartmann}, {Calvet}, {Whitney}, {Barrado},
  {Vrba}, {Covey}, {Herbst}, {Furesz}, {Aigrain}, \&
  {Favata}}]{Staufferetal2014}
{Stauffer}, J., {Cody}, A.~M., {Baglin}, A., {et~al.} 2014, \aj, 147, 83

\bibitem[{{Strom} {et~al.}(1989){Strom}, {Strom}, {Edwards}, {Cabrit}, \&
  {Skrutskie}}]{Strometal1989}
{Strom}, K.~M., {Strom}, S.~E., {Edwards}, S., {Cabrit}, S., \& {Skrutskie},
  M.~F. 1989, \aj, 97, 1451

\bibitem[{{Tomczak} \& {Ciborski}(2007)}]{Tomczak&Ciborski2007}
{Tomczak}, M., \& {Ciborski}, T. 2007, \aap, 461, 315

\bibitem[{{Trigilio} {et~al.}(1993){Trigilio}, {Umana}, \&
  {Migenes}}]{Trigilioetal1993}
{Trigilio}, C., {Umana}, G., \& {Migenes}, V. 1993, \mnras, 260, 903

\bibitem[{{Venuti} {et~al.}(2014){Venuti}, {Bouvier}, {Flaccomio}, {Alencar},
  {Irwin}, {Stauffer}, {Cody}, {Teixeira}, {Sousa}, {Micela}, {Cuillandre}, \&
  {Peres}}]{Venutietal2014}
{Venuti}, L., {Bouvier}, J., {Flaccomio}, E., {et~al.} 2014, \aap, 570, A82

\bibitem[{{Williams} \& {Best}(2014)}]{Williams&Best2014}
{Williams}, J.~P., \& {Best}, W.~M.~J. 2014, \apj, 788, 59

\end{thebibliography}
\end{document}